\documentclass[letterpaper]{JHEP3}
\usepackage{bm}
\usepackage{amsmath}
\usepackage{yfonts}
\usepackage{epsfig,latexsym}
\newcommand{\qn}{\textswab{q}}
\newcommand{\wn}{\textswab{w}}
\newcommand{\<}{\langle}
\renewcommand{\>}{\rangle}
\renewcommand{\d}{\partial}
\newcommand{\N}{{\cal N}}

\renewcommand{\Re}{\mathrm{Re}\,}
\renewcommand{\Im}{\mathrm{Im}\,}

\newcommand{\beq}{\begin{equation}}
\newcommand{\eeq}{\end{equation}}
\newcommand{\bqa}{\begin{eqnarray}}
\newcommand{\eqa}{\end{eqnarray}}
\newcommand{\eden}{\varepsilon}

\title{Relativistic viscous hydrodynamics,\\
 conformal invariance, 
 and holography}

\author{Rudolf Baier \\
Fakult\"at f\"ur Physik, Universit\"at Bielefeld, D-33501 Bielefeld, Germany\\
E-mail: \email{baier@physik.uni-bielefeld.de}
}
\author{Paul Romatschke and Dam Thanh Son \\
Institute for Nuclear Theory, University of Washington, \\
Box 351550, Seattle, WA, 98195, USA\\
E-mail: \email{paulrom@phys.washington.edu}, \email{son@phys.washington.edu}
}

\author{Andrei O.~Starinets \\
School of Physics \& Astronomy, University of Southampton,\\ 
Highfield, 
 Southampton SO17 1BJ, United Kingdom\\
E-mail: \email{starina@phys.soton.ac.uk}
}

\author{Mikhail A.~Stephanov\\
Department of Physics, University of Illinois, Chicago, IL 60607-7059,
USA\\
E-mail: \email{misha@uic.edu}
}

\preprint{BI-TP 2007/29\\INT PUB 07-45\\SHEP-07-47}

\abstract{We consider second-order viscous hydrodynamics in conformal
field theories at finite temperature.  We show that
conformal invariance imposes powerful constraints on the form of the
second-order corrections.  By matching to the AdS/CFT calculations of
correlators, and to recent results for Bjorken flow obtained by Heller
and Janik, we find three (out of five) second-order transport
coefficients in the strongly coupled ${\cal N}=4$ supersymmetric Yang-Mills
theory.  We also discuss how these new coefficents can arise within the
kinetic theory of weakly coupled conformal plasmas.  We point out that
the M\"uller-Israel-Stewart theory, often used in numerical 
simulations, does not contain all allowed second-order terms
and, frequently,
terms required by conformal invariance.}

\begin{document}

\section{Introduction}

Relativistic hydrodynamics is an important theoretical tool in
heavy-ion physics, astrophysics, and cosmology. Hydrodynamics gives
reliable description of the non-equilibrium real-time macroscopic
evolution of a given system. It is an effective description in terms
of a few relevant variables (fields) and it applies to the evolution which is
slow, both in space and in time, relative to a certain microscopic
scale~\cite{Landafshitz6,KadanoffMartin}.

In the most common applications of hydrodynamics the underlying
microscopic theory is a kinetic theory.
In this case the microscopic scale which limits the validity of
hydrodynamics is the mean free path $\ell_{\rm mfp}$. In other words,
the parameter controlling the precision of hydrodynamic approximation is
$k\ell_{\rm mfp}$, where $k$ is the characteristic momentum scale of
the process under consideration.  

More generally, the underlying microscopic description is a quantum
field theory, which might not necessarily admit a kinetic
description. An experimental example of such a system is the strongly
coupled quark-gluon plasma (sQGP) recently discovered at the
Relativistic Heavy-Ion Collider (RHIC) at Brookhaven National
Laboratory.  The ${\cal N}=4$ supersymmetric $SU(N_c)$ Yang-Mills
theory in the limit of strong coupling provides a theoretical example
of such a system which, in the limit of large number of colors $N_c$,
can be studied analytically using the AdS/CFT
correspondence~\cite{Maldacena:1997re}.  In these cases, where kinetic
description may be absent, the role of the parameter $\ell_{\rm mfp}$
is played by some typical microscopic scale. In the above examples
this scale is set by the temperature: $\ell_{\rm mfp}\sim T^{-1}$.


When the parameter $k\ell_{\rm mfp}$ is not too small, one may want to
go beyond the first order in $k\ell_{\rm mfp}$.  This is the case, for
example, in the early stages of heavy-ion collisions.  There are two
sources of corrections beyond the $k\ell_{\rm mfp}$ order.  First,
there are corrections due to thermal fluctuations of hydrodynamic
variables contributing via nonlinearities of the hydrodynamic
equations. The fluctuation corrections lead to nonanalytic
low-momentum behavior of certain correlators~\cite{Kovtun:2003vj}
(similarly to the chiral logarithms that emerge from loops in chiral
perturbation theory) and are, for example, essential for describing
non-trivial dynamical critical behavior near phase
transitions~\cite{Hohenberg:1977ym}.  Such corrections are
calculable in the framework of hydrodynamics with thermal noise.

The second source of corrections are second-order terms (order
$(k\ell_{\rm mfp})^2$) in the hydrodynamic equations, sometimes called
the Burnett corrections \cite{Burnett}. These corrections come with
additional transport coefficients. These second-order transport
coefficients are not calculable from hydrodynamics, but have to be
determined from underlying microscopic description or input
phenomenologically, similarly to first-order transport coefficients
such as viscosity.

In gauge theories with a large number of colors $N_c$ the corrections of
the first type (fluctuation) are suppressed by $1/N_c^2$ ~\cite{Kovtun:2003vj} 
 and therefore the corrections of the second
type (Burnett) dominate in the limit of fixed $k$ and
$N_c\to\infty$.  For this reason, in this paper, we concentrate on the
second type of corrections.  Moreover, we shall consider the case of
conformal theories, where the number of second-order
transport coefficients is substantially reduced.  In the real-world
applications we deal with fluids which are not exactly conformal, 
however, e.g., QCD  at sufficiently high
temperatures is approximately conformal.

This paper is organized as follows.  In Sec.~\ref{sec:Weyl} we derive
the consequences of conformal symmetry for hydrodynamics.  In
Sec.~\ref{sec:2nd} we classify all terms of order $k^2$ consistent
with conformal symmetry.  In Sec.~\ref{sec:SYM} we compute three of
the five new transport coefficients for the strongly-coupled ${\cal
  N}=4$ supersymmetric Yang-Mills (SYM) theory using the AdS/CFT
correspondence.  In Sec. \ref{sec:kinetic-theory} we show
that hydrodynamic equations derived from the kinetic description 
(Boltzmann equation) of
a weakly coupled conformal theory do not contain all allowed
second-order terms. In Sec.~\ref{section:IS}, we analyze our
findings from the point of view of the M\"uller-Israel-Stewart theory
\cite{Mueller1,Israel:1976tn,IS0b,Israel:1979wp}, which involves only one new parameter
at the second order, and show that this parameter cannot account for all
second-order corrections.  Our conclusions are summarized in
Sec.~\ref{sec:concl}.

\section{Conformal invariance in hydrodynamics}
\label{sec:Weyl}

To set the stage, let us emphasize again that
hydrodynamics is a controlled expansion scheme ordered by the power of
the parameter $k\ell_{\rm mfp}$, or equivalently, by the number of
derivatives of the hydrodynamic fields. We shall set up this expansion
paying particular attention to the consequences of the conformal
invariance on the equations of hydrodynamics.

\subsection{Conformal invariance and Weyl anomalies}
\label{sec:weyl1}

The hydrodynamic fields are expectation values of certain quantum
fields, such as e.g., components of the stress-energy tensor, averaged
over small but macroscopic volumes and time intervals. Such averages
can, in principle, be calculated in the close-time-path (CTP)
formalism \cite{CTP}.  Consider a generic finite-temperature field
theory in the CTP formulation.
Turning on external metrics on the
upper and lower contours, the partition function is
\begin{equation}\label{Z-CPT}
  Z[g_{\mu\nu}^1, g_{\mu\nu}^2] = \int\!{\cal D}\phi_1\,{\cal D}\phi_2\, 
  \exp\left\{iS[\phi_1,g_{\mu\nu}^1]-iS[\phi_2,g_{\mu\nu}^2]\right\} ,
\end{equation}
where $\phi_1$ and $\phi_2$ represent the two sets of all fields 
living on the upper and lower parts of the contours, and $S[\phi,g_{\mu\nu}]$
is the general coordinate invariant action.

The one-point Green's function of the stress-energy tensor is obtained
by differentiating the partition function (the metric signature here
is $-+++$):
\begin{align}\label{T12}
  \< T^{1\mu\nu} \> &= -\frac{2i}{\sqrt{-g_1}}
   \frac{\delta\ln Z}{\delta g_{\mu\nu}^1}\,,\\
  \< T^{2\mu\nu} \> &= \phantom{+}\frac{2i}{\sqrt{-g_2}}
  \frac{\delta\ln Z}{\delta g_{\mu\nu}^2}\,,
\end{align}
where $\<\ldots\>$ denote the mean value under the path integral and 
$\sqrt{-g_{1,2}}\equiv\sqrt{-{\rm det} g_{\mu \nu}^{1,2}}$.

In this paper we consider conformally invariant theories.  In such
theories the action $S[\phi,g_{\mu\nu}]$ evaluated on classical
equations of motion $\delta S/\delta\phi=0$ and viewed as a functional
of the external metric $g_{\mu\nu}$ is invariant under local
dilatations, or Weyl transformations:
\begin{equation}
  \label{eq:Weyl-gmunu}
  g_{\mu\nu}\to e^{-2\omega} g_{\mu\nu}, 
\end{equation}
with parameter $\omega$ a function of space-time coordinates. As a consequence,
classical stress-energy tensor $T^{\mu\nu}_{\rm cl}\equiv \delta S/ \delta g_{\mu\nu}$
is traceless since $g_{\mu\nu}T^{\mu\nu}_{\rm cl}=-(1/2)\delta S/\delta\omega=0$.

In the conformal quantum theory (\ref{Z-CPT}) the Weyl anomaly 
\cite{Birrell:1982ix,Duff:1993wm} implies
\begin{subequations}\label{Weyl}
\begin{align}
  g^1_{\mu\nu} \< T^{1\mu\nu}\> &= W_d[g^1_{\mu\nu}],\\
  g^2_{\mu\nu} \< T^{2\mu\nu}\> &= W_d[g^2_{\mu\nu}],
\end{align}
\end{subequations}
where $W_d$ is the Weyl anomaly in $d$ dimensions, which is identically zero
for odd $d$.  For $d=4$:
\begin{equation}
  W_4[g_{\mu\nu}]=  -\frac 
  a{16\pi^2}(R_{\mu\nu\lambda\rho}R^{\mu\nu\lambda\rho}
  - 4 R_{\mu\nu} R^{\mu\nu} + R^2)
  + \frac c{16\pi^2}
  (R_{\mu\nu\lambda\rho}R^{\mu\nu\lambda\rho}
  - 2 R_{\mu\nu} R^{\mu\nu} + \tfrac13 R^2),
\end{equation}
where $R_{\mu \nu \lambda \rho}$ and $R_{\mu \nu}$ ($R$) 
are the Riemann tensor and Ricci tensor (scalar),
and for $SU(N_c)$ ${\mathcal N}=4$ SYM
theory $a=c=\frac{1}{4}\left(N_c^2-1\right)$  \cite{Aharony:1999ti}.
The right-hand side of Eqs.~(\ref{Weyl}) contains four derivatives.
In general, for even $d=2k$, $W_{2k}$ contains $2k$ derivatives of the
metric.

Let us now explore the consequences of Weyl anomalies for
hydrodynamics.  The hydrodynamic equations (without noise) do not
capture the whole set of CTP Green's functions, but only the retarded
ones.  Hydrodynamics determines the stress-energy tensor $T^{\mu\nu}$
(more precisely, its slowly varying average over sufficiently long
scales) in the presence of an arbitrary (also slowly varying)
source $g_{\mu\nu}$.  The connection to
the CTP partition function can be made explicit by writing
\begin{equation}\label{ggamma}
  g^1_{\mu\nu} = g_{\mu\nu} + \frac12\gamma_{\mu\nu},\qquad
  g^2_{\mu\nu} = g_{\mu\nu} - \frac12\gamma_{\mu\nu}.
\end{equation}
If $\gamma_{\mu\nu}=0$ then $Z=1$ since the time evolution on the lower
contour exactly cancels out the time evolution on the upper contour.
When $\gamma_{\mu\nu}$ is small one can expand
\begin{equation}\label{lnZgamma}
  \ln Z = \frac i 2\int\!dx\, \sqrt{-g(x)}\, \gamma_{\mu\nu}(x)
  T^{\mu\nu}(x) + O(\gamma^2),
\end{equation}
where $T^{\mu\nu}(x)$ depends on $g_{\mu\nu}$, and is the
stress-energy tensor in the presence of the source $g_{\mu\nu}$.  At
long distance scales it should be the same as computed from hydrodynamics.

Substituting Eqs.~(\ref{ggamma}) and (\ref{lnZgamma}) into
Eq.~(\ref{Weyl}), the $O(1)$ and $O(\gamma)$ terms yield two
equations:
\begin{subequations}\label{Weyl12}
\begin{align}
   g_{\mu\nu} T^{\mu\nu} &= W_d[g_{\mu\nu}],\\
  g_{\mu\nu}(x) \frac{\delta[\sqrt{-g(x)}\, T^{\alpha \beta}(x)]}
  {\delta g_{\mu \nu}(y)} + \sqrt{-g(x)} T^{\alpha\beta}(x)\delta^d(x-y)&=
  \frac\delta{\delta g_{\alpha\beta}(y)} (\sqrt{-g(x)}\, W_d[g_{\mu\nu}(x)]).
\end{align}
\end{subequations}
In odd dimensions, the right hand sides of Eqs.~(\ref{Weyl12}) are
zero.  In even dimensions, they contain $d$ derivatives.  In a
hydrodynamic theory, where one keeps less than $d$ derivatives, they
can be set to zero.  For example, at $d=4$, the Weyl anomaly is
visible in hydrodynamics only if one keeps terms to the fourth order in
derivatives.  This is two orders higher than in second-order
hydrodynamics considered in this paper.  For larger even $d$, one has
to go to even higher orders to see the Weyl anomaly.  Thus, we can
neglect $W_d$ on the right hand side: second-order hydrodynamic theory
is invariant under Weyl transformations.  The two
conditions~(\ref{Weyl12}) then become
\begin{align}
  g_{\mu\nu} T^{\mu\nu} & = 0,\\
  g_{\mu\nu} \frac{\delta T^{\alpha\beta}(y)}
    {\delta g_{\mu\nu}(x)} &= 
  - \left(\frac d2 +1\right) \delta^d(x-y)T^{\alpha\beta}(x). \label{Weyl3} 
\end{align}
Since the r.h.s. of equation~(\ref{Weyl3}) is $-(1/2)\delta T^{\mu\nu}/\delta\omega$ it implies the following tranformation law for 
$T^{\mu\nu}$ under Weyl transformations (\ref{eq:Weyl-gmunu}):
\begin{equation}\label{Tconf}
  T^{\mu\nu}\to e^{(d+2)\omega}\, T^{\mu\nu}.
\end{equation}
Noting that $\ln Z$ is invariant under Weyl transformations
this could have been gleaned from Eq.~(\ref{lnZgamma}) already.

A simple rule of thumb is that for tensors transforming
homogeneously 
\begin{equation}
  \label{eq:A-transform}
 A^{\mu_1\ldots\mu_m}_{\nu_1\ldots\nu_n}\to e^{\Delta_A\,\omega}A^{\mu_1\ldots\mu_m}_{\nu_1\ldots\nu_n},  
\end{equation}
the conformal weight
$\Delta_A$ equals the mass dimension plus the difference between the
number of contravariant and covariant indices:
\begin{equation}
  \label{eq:confdimension}
  \Delta_A = [A] + m - n.
\end{equation}

\subsection{First order hydrodynamics as derivative expansion}
\label{sec:1st-order-hydro}

The existence of hydrodynamic description owes itself to the presence
of conserved quantities, whose densities can evolve (oscillate or
relax to equilibrium) at arbitrarily long times provided the
fluctuations are of large spatial size. Correspondingly, the 
expectation values of such densities are the hydrodynamic
fields. 

In the simplest case we shall consider here, i.e., in a theory without
conserved charges, there are 4 such hydrodynamic fields: energy
density $T^{00}$ and 3 components of the momentum density $T^{0i}$. 
It is common and convenient to
use the local velocity $u^\mu$ instead of the momentum density
variable. It can be defined as the boost velocity needed to go
from the local rest frame, where the momentum density $T^{0i}$
vanishes, back to the lab frame. Similarly, it is convenient to use
$\eden$ -- the energy density in the local rest frame -- instead of
the $T^{00}$ in the lab frame.
The 4 equations for thus defined variables $\eden$ and $u^\mu$
are conservation equations of the energy-momentum tensor
$\nabla_\mu T^{\mu\nu}=0$.

In a covariant form the above definitions of $\eden$ and $u^\mu$ 
can be summarized as
\begin{equation}
  \label{eq:Tmunu-0}
  T^{\mu\nu}=\eden\, u^\mu u^\nu + T^{\mu\nu}_\perp .
\end{equation}
In hydrodynamics, the remaining components $T^{\mu\nu}_\perp$ (spatial
in the local rest frame: $u_\mu T^{\mu\nu}_\perp=0$) 
of the stress-energy tensor $T^{\mu\nu}$
appearing in the conservation equations are not independent variables,
but rather instantaneous functions of the hydrodynamic variables
$\eden$ and $u^\mu$ and their derivatives. In the hydrodynamic
limit, this is the consequence of the fact that the hydrodynamic modes
are infinitely slower than all other modes, the latter therefore can
be integrated out. All quantities appearing in hydrodynamic equations
are averaged over these fast modes, and are functions of the slow
varying hydrodynamic variables.  The functional dependence of
$T^{\mu\nu}_\perp$ (constituitive equations) can be expanded in
powers of 
derivatives of $\eden$ and $u^\mu$.

Writing the most general form
of this expansion consistent with symmetries gives, up to 1st order in
derivatives,
\begin{equation}\label{eq:tmunu-1st-order}
  T^{\mu\nu}_\perp=P(\eden)\Delta^{\mu\nu} - 
\eta(\eden) \sigma^{\mu\nu}
- \zeta(\eden) \Delta^{\mu\nu} 
     (\nabla {\cdot} u) ,
\end{equation}
where the symmetric, transverse tensor with no derivatives
$\Delta^{\mu\nu}$ is given by
\begin{equation}
  \label{eq:Deltamunu}
  \Delta^{\mu\nu}=g^{\mu\nu}+u^\mu u^\nu\,.
\end{equation}
In the local rest frame it is the projector on the spatial subspace.
The symmetric, transverse and traceless 
tensor of first derivatives
$\sigma^{\mu\nu}$ is defined by
\begin{equation}\label{sigmamunu}
  \sigma^{\mu\nu} = 2{}^\<\nabla^{\mu} u^{\nu}{}^\>\, , 
\end{equation}
where for a second rank tensor $A^{\mu\nu}$ the tensor defined as
\begin{equation}
  {}^\<A^{\mu\nu}{}^\>
 \equiv \frac12 \Delta^{\mu\alpha} \Delta^{\nu\beta}
     (A_{\alpha\beta} + A_{\beta\alpha}) 
  - \frac1{d-1} \Delta^{\mu\nu} \Delta^{\alpha\beta} A_{\alpha\beta}\, 
\equiv A^{\<\mu\nu\>}
\end{equation}
is transverse $u_\mu A^{\<\mu\nu\>} = 0$ (i.e., only spatial components
in the local rest frame are nonzero) and traceless $g_{\mu\nu}
A^{\<\mu\nu\>}=0$.

In the gradient expansion (\ref{eq:tmunu-1st-order}), the scalar function
$P(\eden)$ can be identified as the thermodynamic 
pressure (in equilibrium, when all the gradients vanish),
while $\eta(\eden)$ and $\zeta(\eden)$ are the shear and bulk viscosities.
The expansion coefficients $P$, $\eta$ and $\zeta$ 
are determined by the physics of the fast
(non-hydrodynamic, microscopic)
modes that have been integrated out.

\subsection{Conformal invariance in first-order hydrodynamics}


It is straightforward to check that if $T^{\mu\nu}$ transforms as in
Eq.~(\ref{Tconf}) and $T^\mu_\mu=0$, then its covariant divergence
transforms homogeneously: $\nabla_\mu T^{\mu\nu} \to
e^{(d+2)\omega}\nabla_\mu T^{\mu\nu}$, hence the hydrodynamic equation
$\nabla_\mu T^{\mu\nu}=0$ is Weyl invariant~\cite{Bhattacharyya:2007vs}.

Let us now see what restrictions conformal invariance imposes on 
the first-order constitutive equations (\ref{eq:tmunu-1st-order}).  
First, the tracelessness condition $T^\mu_\mu=0$ forces $\eden=(d-1)P$ and $\zeta=0$. 
Since in a conformal theory $\eden={\rm const} {\cdot} T^{d}$, we
shall trade $\eden$ variable for $T$ in what follows.
Since $g_{\mu\nu}u^\mu u^\nu = -1$ the conformal weight of $u^\mu$ is 1.
By definition (\ref{eq:Tmunu-0}) and by (\ref{Tconf})
  $\eden$ has conformal weight
$d$ and therefore
\begin{equation}\label{Tuconf}
  T \to e^\omega T, \qquad u^\mu \to e^\omega u^\mu
\end{equation}
in accordance with the simple rule (\ref{eq:confdimension}).

By direct computation we find that
\begin{equation}
  \sigma^{\mu\nu} \to e^{3\omega} \sigma^{\mu\nu},
\end{equation}
i.e. $\sigma^{\mu\nu}$ transforms homogeneously with conformal weight 3 independent of $d$ 
(in agreement with (\ref{eq:confdimension})).
For conformal fluids $\eta={\rm const}\cdot T^{d-1}$, and therefore $T^{\mu\nu}$ transforms
homogeneously under Weyl transformation as in Eq.~(\ref{Tconf}).

\section{Second-order hydrodynamics of a conformal fluid}
\label{sec:2nd}

In this Section we shall continue the derivative expansion
(\ref{eq:tmunu-1st-order}). We shall write down all possible
second-order terms in the stress-energy tensor allowed by Weyl
invariance.  Then
 we shall compute the coefficients in front of these
terms in the ${\cal N}=4$ SYM plasma by matching hydrodynamic correlation
functions  with gravity calculations in Section~\ref{sec:SYM}.

\subsection{Second-order terms}
\label{sec:2nd-terms}

Rewriting Eq.~(\ref{eq:Tmunu-0}) 
we introduce the dissipative part of the stress-energy tensor,
$\Pi^{\mu\nu}$:
\begin{equation}\label{eq:Pimunu-def}
  T^{\mu\nu} = \eden u^\mu u^\nu + P \Delta^{\mu\nu} + \Pi^{\mu\nu}\,,
\end{equation}
which contains only the derivatives and vanishes
in a homogeneous equilibrium state. The tensor $\Pi^{\mu\nu}$ is
symmetric and transverse, $u_\mu\Pi^{\mu\nu}=0$. For
conformal fluids it must be also traceless $g_{\mu\nu}\Pi^{\mu\nu}=0$.
To first order
\begin{equation}
  \label{eq:Pimunu-1st-order}
  \Pi^{\mu\nu}=-\eta\sigma^{\mu\nu} + \textrm{(2nd order terms)}, 
\end{equation}
where $\sigma^{\mu\nu}$ is defined in Eq.~(\ref{sigmamunu}).
We will also use the notation for the vorticity
\begin{equation}
  \Omega^{\mu\nu} = 
  \frac12   
    \Delta^{\mu\alpha}\Delta^{\nu\beta}
    (\nabla_\alpha u_\beta - \nabla_\beta u_\alpha)\, .
\end{equation}

We note that in writing down second-order terms in $\Pi^{\mu\nu}$, one
can always rewrite 
the derivatives along the $d$-velocity direction 
\begin{equation}
  \label{eq:D-deriv}
  D\equiv
u^\mu\nabla_\mu
\end{equation}
 (temporal derivative in the local rest frame)
in terms of transverse (spatial in the local rest frame) derivatives through the
zeroth-order equations of motion:
\begin{equation}\label{0ord}
  D \ln T = - \frac1{d-1} (\nabla_\perp \cdot u),\quad
  D u^\mu = - \nabla_\perp^\mu \ln T, \qquad
  \nabla^\mu_\perp \equiv \Delta^{\mu\alpha}\nabla_\alpha\, .
\end{equation}
Notice also that $\nabla_\perp\!\cdot u = \nabla\cdot u$.

With the restriction of transversality and tracelessness, there are eight 
possible contributions to the stress-energy tensor:
\begin{equation}
\begin{split}
  &\nabla^{\<\mu}\ln T\, \nabla^{\nu\>}\ln T, \quad
  \nabla^{\<\mu}\nabla^{\nu\>} \ln T, \quad
  \sigma^{\mu\nu}(\nabla{\cdot}u), \quad
  {\sigma^{\<\mu}}_\lambda \sigma^{\nu\>\lambda}\\
  & {\sigma^{\<\mu}}_\lambda \Omega^{\nu\>\lambda}, \quad
  {\Omega^{\<\mu}}_\lambda \Omega^{\nu\>\lambda}, \quad
  u_\alpha R^{\alpha\<\mu\nu\>\beta} u_\beta,\quad
  R^{\<\mu\nu\>}\, .
\end{split}
\end{equation}

\newcommand\op[1]{{\cal O}_{#1}^{\mu\nu}}

By direct computations we find that there are only five combinations
that transform homogeneously under Weyl tranformations.  They are
\begin{align}
  & \op{1}= R^{\<\mu\nu\>} - (d-2) \left(\nabla^{\<\mu} \nabla^{\nu\>} \ln T 
    - \nabla^{\<\mu} \ln T\, \nabla^{\nu\>} \ln T\right),
  \label{Rmunu}\\
  & \op{2}= R^{\<\mu\nu\>}-(d-2) u_\alpha R^{\alpha\<\mu\nu\>\beta} u_\beta\, ,\\
  & \op{3}={\sigma^{\<\mu}}_\lambda \sigma^{\nu\>\lambda}\,, \qquad
  \op{4}= {\sigma^{\<\mu}}_\lambda \Omega^{\nu\>\lambda}\,, \qquad
   \op{5}={\Omega^{\<\mu}}_\lambda \Omega^{\nu\>\lambda} \,.
\end{align}

In the linearized hydrodynamics in flat space only the term
$\op{1}$ contributes.
For convenience and to facilitate the comparision with the
Israel-Stewart theory we shall use instead of (\ref{Rmunu}) 
the term
\begin{equation}\label{Dsigma}
 {}^\< D \sigma^{\mu\nu}{}^\>+ \frac 1{d-1}\sigma^{\mu\nu}(\nabla{\cdot} u)
\end{equation}
which, with (\ref{0ord}), reduces to the linear combination: 
$\op{1}-\op{2}-(1/2)\op{3}-2\op{5}$.
It is straightforward to check directly that~(\ref{Dsigma}) transforms
homogeneously under Weyl transformations.

Thus, our final expression for the dissipative part of the
stress-energy tensor, up to second order in derivatives, is
\begin{equation}\label{T2nd}
\begin{split}
  \Pi^{\mu\nu} & = -\eta \sigma^{\mu\nu}\\
   &\quad+ \eta\tau_\Pi \left[ {}^\<D\sigma^{\mu\nu}{}^\> + \frac1{d-1}\sigma^{\mu\nu}
    (\nabla{\cdot}u) \right]
  +\kappa\left[R^{\<\mu\nu\>}-(d-2) u_\alpha R^{\alpha\<\mu\nu\>\beta} 
     u_\beta\right]\\
  &\quad + \lambda_1 {\sigma^{\<\mu}}_\lambda \sigma^{\nu\>\lambda}
  + \lambda_2 {\sigma^{\<\mu}}_\lambda \Omega^{\nu\>\lambda}
  + \lambda_3 {\Omega^{\<\mu}}_\lambda \Omega^{\nu\>\lambda}\, .
\end{split}
\end{equation}
The five new constants are $\tau_\Pi$, $\kappa$, $\lambda_{1,2,3}$.  
Note that using lowest order relations $\Pi^{\mu\nu}=-\eta\sigma^{\mu\nu}$,
Eqs.(\ref{0ord}) and 
$D\eta=-\eta\,\nabla{\cdot}u $, Eq.~(\ref{T2nd})  may be rewritten in the form
\begin{equation}\label{ISlikePi}
\begin{split}
\Pi^{\mu\nu} & = -\eta \sigma^{\mu\nu}- \tau_\Pi \left[ {}^\<D\Pi^{\mu\nu}{}^\> 
 + \frac d{d-1} \Pi^{\mu\nu}
    (\nabla{\cdot}u) \right] \\
  &\quad 
  + \kappa\left[R^{\<\mu\nu\>}-(d-2) u_\alpha R^{\alpha\<\mu\nu\>\beta} 
      u_\beta\right]\\
  & + \frac{\lambda_1}{\eta^2} {\Pi^{\<\mu}}_\lambda \Pi^{\nu\>\lambda}
  - \frac{\lambda_2}\eta {\Pi^{\<\mu}}_\lambda \Omega^{\nu\>\lambda}
  + \lambda_3 {\Omega^{\<\mu}}_\lambda \Omega^{\nu\>\lambda}\, .
\end{split}
\end{equation}
This equation is, in form, similar to an equation of the Israel-Stewart
theory (see Section~\ref{section:IS}).  
In the linear regime it actually coincides with the Israel-Stewart
theory (\ref{eq:DPimunu}).
We emphasize, however, that one cannot claim that 
Eq.~(\ref{ISlikePi}) captures all orders in the momentum expansion
(see Section \ref{section:IS}).

Further remarks are in order.  First, the $\kappa$ term vanishes
in flat space.  If one is interested in solving the hydrodynamic
equation in flat space, then $\kappa$ is not needed.  Nevertheless,
$\kappa$ contributes to the two-point Green's function of the
stress-energy tensor.  We emphasize that the term proportional to $\kappa$
is not a contact term, since it contains $u^\mu$.
The $\lambda_{1,2,3}$ terms are
nonlinear in velocity, so are not needed if one is looking at small
perturbations (like sound waves).  For irrotational flows
$\lambda_{2,3}$ are not needed.  The parameter $\tau_\Pi$ has
dimension of time and can be thought of as the relaxation time.  This
interpretation of $\tau_\Pi$ can be most clearly seen from 
Eq.~(\ref{ISlikePi}). For further discussion, see Section~\ref{section:IS}.

\subsection{Kubo's formulas}
\label{kf}


To relate the new kinetic coefficients with thermal correlators,
first let us consider the
response of the fluid to small and smooth metric perturbations.  We shall
moreover restrict ourselves to a particular type of perturbations which
is simplest to treat using AdS/CFT correspondence.  Namely, for dimensions
$d\ge4$ we take $h_{xy}=h_{xy}(t,z)$.
 For $d=3$, there are only two spatial coordinates,
so we take $h_{xy}=h_{xy}(t)$.
Since it is a tensor perturbation the fluid
remains at rest: $T=\textrm{const}$, $u^\mu=(1,{\bf 0})$.  Inserting this
into Eq.~(\ref{T2nd}) we find, for $d\ge4$,
\begin{equation}
  T^{xy}= -P h_{xy} -\eta \dot h_{xy} + \eta\tau_\Pi \ddot h_{xy}
          -\frac\kappa 2 [(d-3)\ddot h_{xy} + \d_z^2 h_{xy}]\, .
\end{equation}
The linear response theory implies that the retarded Green's function
in the tensor channel is
\begin{equation}
  G_R^{xy,xy}(\omega, k) = P - i\eta\omega + \eta\tau_\Pi \omega^2
      -\frac\kappa 2 [(d-3)\omega^2+k^2]\, .
\label{hydroxyxycorr}
\end{equation}
For $d=3$ there is no momentum $k$, and the formula becomes
\begin{equation}
  G_R^{xy,xy}(\omega) = P - i\eta\omega + \eta\tau_\Pi \omega^2,
  \qquad d=3\, .     
\label{hydroxyxycorr3}
\end{equation}

Thus the two kinetic coefficients $\tau_\Pi$ and $\kappa$ can be found
from the coefficients of the $\omega^2$ and $k^2$ terms in the
low-momentum expansion of $G_R^{xy,xy}(\omega,k)$ in the case of
$d\ge4$, and just from the $\omega^2$ term in the case of $d=3$.


\subsection{Sound Pole}
\label{ssp}

We now turn to another way to determine $\tau_\Pi$, which is based on
the position of the sound pole.  The fact that we have two independent
methods to determine $\tau_\Pi$ allows us to check the
self-consistency of the calculations.


To obtain the dispersion relation, we consider a (conformal)
hydrodynamic system in stationary equilibrium, that is, with fluid
velocity $u^\mu=(1,{\bf 0})$, homogeneous energy density
$\eden={\rm const} \cdot T^d$ and $\Pi^{\mu\nu}=0$. The
speed of sound is defined by $c_s^2=d P(\eden)/d\eden$.
In conformal theory it is a constant: $c_s^2=1/(d-1)$.  Now let us slightly
perturb the system and denote the departure from equilibrium
energy density, velocity, and stress as $\delta \eden$, $u^i$, and~$\Pi^{ij}$.

For small perturbations, one can neglect the nonlinear terms in
Eq.~(\ref{ISlikePi}) and the hydrodynamic equations are identical to
those of the Israel-Stewart theory.  For completeness, we rederive
here the sound dispersion in this theory.  
To linear approximation in the perturbations, we have
\beq
\delta T^{00}=\delta \eden,\quad \delta T^{0i}=(\eden+P) u^i,
\quad \delta T^{ij}=c_s^2 \delta \eden\ \delta^{ij}+\Pi^{ij}.
\eeq

For sound waves travelling in $x$ direction 
we take $u^x$ and $\Pi^{xx}$ as the only nonzero
components of $u^i$ and $\Pi^{ij}$, and dependent only on $x$ and $t$.  
Energy-momentum conservation implies
\begin{align}\label{eq:deltaepsilon}
\partial_0 (\delta\eden)+(\eden+P)\partial_x u^x &=0\, ,\\
\label{eq:deltau}
(\eden+P) \partial_0 u^x+c_s^2\partial_x (\delta\eden)+
\partial_x \Pi^{x x}&=0\, .
\end{align}
Eq.~(\ref{ISlikePi}) has the form
\begin{equation}\label{eq:deltaPi}
\tau_\Pi \partial_0 \Pi^{xx}+\Pi^{xx}=
- \frac{2(d-2)}{d-1}\eta \partial_x u^x\, .
\end{equation}

For a plane wave, equations (\ref{eq:deltaepsilon}), (\ref{eq:deltau}) and
(\ref{eq:deltaPi}) give the dispersion relation
\begin{equation}
  - \omega^3 \tau_\Pi- i \omega^2+\omega k^2 c_s^2 \tau_\Pi
  +\omega k^2 \frac{2(d-2)}{d-1}\frac{\eta}{\eden+P}+i k^2 c_s^2=0.
\end{equation}
At small $k$, the two solutions of this equation corresponding to the
sound wave are
\begin{equation}\label{dispsound1}
  \omega_{1,2} = \pm c_s k - i\Gamma k^2 \pm \frac\Gamma{c_s} 
  \left( c_s^2\tau_\Pi - \frac\Gamma2\right) k^3
+{\cal O}(k^4)\,,
\end{equation}
where
\begin{equation}
\Gamma = \frac{d-2}{d-1}\,
  \frac\eta{\eden+P}\,.
\end{equation}
The third solution is given by
\begin{equation}
  \label{eq:sound-omega3}
  \omega_3=-i\tau_\Pi^{-1}+O(k^2)\,.
\end{equation}
 Since $\omega_3$ does not vanish as $k\to0$,
  but remains on the order of a macroscopic scale, this third solution
lies beyond the regime of validity of hydrodynamics (see also
discussion in Section~\ref{section:IS}).

\subsection{Shear pole}
\label{sec:shear-pole}

 In hydrodynamics, there exists an overdamped mode describing fluid flow 
in a direction perpendicular to the velocity gradient, e.g., with $u_y\sim
e^{-i\omega t+ikx}$.  First-order hydrodynamics gives the
leading-order dispersion relation, $\omega=-i\eta k^2/(\eden+P)$.
The next correction to this dispersion relation is proportional to $k^4$ and thus is beyond
the reach of the second-order theory.  This correction can be fully   
determined only in third-order hydrodynamics. To illustrate that, we
shall compute this correction here, taking the second-order theory
literally and pretending the third-order terms are not contributing.  
We shall than find the  expected mismatch between this (incorrect) result and
the AdS/CFT computation in the strongly coupled ${\cal
N}=4$ SYM theory.

The perturbation corresponding to the fluid flowing in the $y$ direction
with velocity gradient along the $x$ direction (shear flow) involves the variables
\beq
  u^y(t,x), \qquad \Pi^{xy}(t,x)\, ,
\eeq
such that we get from $\partial_\mu \delta T^{\mu \nu}=0$
\bqa
(\eden+P) \partial_0 u^y+\partial_x \Pi^{x y}&=&0.
\label{anothereq}
\eqa
From 
Eq.~(\ref{ISlikePi}) we find
\begin{equation}\label{eq:shear-Pixy}
\tau_\Pi \partial_0 \Pi^{xy}+\Pi^{xy}=
- \eta \partial_x u^y.
\end{equation}
The dispersion relation is determined by
\begin{equation}\label{dispsheareq}
  - \omega^2 \tau_\Pi- i \omega+ k^2 \frac{\eta}{\eden+P} =0
\end{equation}
so the shear mode dispersion relation in the limit $k\rightarrow 0$ becomes
\begin{equation}
  \omega = -i h k^2-i h^2 {\tau_\Pi} k^4+\mathcal{O}(k^6) ,
  \qquad h = \frac\eta{\eden+P}\,.
  \label{dispshear1}
\end{equation}
The second solution, $\omega=-i\tau_\Pi^{-1} +O(k^2)$, is obviously
beyond the regime of validity of the hydrodynamic equation (see also
Section~\ref{section:IS}).

It is easy to see that expression~(\ref{dispshear1}) 
unjustifiably exceeds the precision of
 the second-order theory: the kept correction is ${\cal O}(k^2)$
relative to the leading-order term, instead of being~$O(k)$.  We can
trace this to Eq.~(\ref{dispsheareq}), in which we keep terms to
second order in $\omega$ and~$k$.  For shear modes, however,
$\omega\sim k^2$, and the term $\omega^2$ that we keep in
Eq.~(\ref{dispsheareq}) is of the same order of magnitude as terms
$O(k^4)$ omitted in Eq.~(\ref{dispsheareq}).  
The latter term can appear if the equation~(\ref{eq:shear-Pixy}) 
for $\Pi^{xy}$
contains a term $\d_x^3 u^y$ that may appear in third-order
hydrodynamics.  This is beyond the scope of this paper.

\subsection{Bjorken Flow}
\label{bf}

So far, we have studied only quantities involved in the linear
response of the fluid, for which linearized hydrodynamics suffices.
In order to determine the coefficients $\lambda_{1,2,3}$, one must
consider nonlinear solutions to the hydrodynamic equations.  One  
 such solution is the Bjorken boost-invariant flow
\cite{Bjorken:1982qr}, relevant to relativistic heavy-ion collisions.

Since hydrodynamic equations are boost-invariant, a solution with
boost-invariant initial conditions will remain boost invariant.  The
motion in the Bjorken flow is a one-dimensional expansion, 
along an axis which we
choose to be $z$, with local velocity equal to $z/t$. 
The most convenient are the comoving coordinates: proper time 
for each local element 
$\tau=\sqrt{t^2-z^2}$ and rapidity $\xi={\rm arctanh}\,(z/t)$.
In these coordinates each element is at rest:
$(u^\tau,u^\xi,\bm u^\perp)=(1,0,\bm 0)$. 

The motion is irrotational, and thus we can only determine the coefficient
$\lambda_1$, but not $\lambda_2$ or $\lambda_3$.

Since velocity $u^\mu$ is constant in the coordinates we chose, 
the only nontrivial equation
is the equation for the energy density:
\begin{equation}
  \label{eq:eden-eqn-bj}
  D\eden + (\eden + P)\nabla\cdot u + \Pi^{\mu\nu}\nabla_\mu u_\nu = 0.
\end{equation}
Boost invariance means that $\eden(\tau)$ is a function of $\tau$
only.  The metric is given by
$  ds^2 = -d\tau^2 + \tau^2 d\xi^2 + d\bm x_\perp^2$
and it is easy to see
that the only nonzero component of $\nabla_\mu u_\nu$ is $\nabla_\xi
u_\xi=\tau$. Using $P=\eden/(d-1)$ we can write:
\begin{equation}
  \label{eq:eden-eqn-bj-2}
\partial_\tau \eden + \frac{d}{d{-}1}\,\frac\eden\tau
=
  -\tau\, \Pi^{\xi\xi}.
\end{equation}

For large $\tau$, the viscous contribution on the r.h.s. in
(\ref{eq:eden-eqn-bj-2}) becomes negligible and the asymptotics of the
solution is thus given by
\begin{equation}
  \label{eq:eden-ideal-asymp}
  \eden(\tau) = C\,\tau^{-2+\nu}+\textrm{(viscous corrections)},
\quad \textrm{where}\quad \nu\equiv \frac{d-2}{d-1},
\end{equation}
and $C$ is the integration constant.
As we shall see, the expansion parameter in
(\ref{eq:eden-ideal-asymp}) is $\tau^{-\nu}$.

Calculating the r.h.s. of Eq.~(\ref{eq:eden-eqn-bj-2}) using
Eq.~(\ref{T2nd}) we find
\begin{equation}
  \label{eq:Pi-xixi}
  - \tau\, \Pi^{\xi\xi} = 
 2 \nu \eta \tau^{-2}
+ 2\nu^2 \left(\eta \tau_\Pi - 2\lambda_1\,\frac{d{-}3}{d{-}2}\right)\tau^{-3}
+ \ldots
\,.
\end{equation}
Integrating equation~(\ref{eq:eden-eqn-bj-2}), one should take into
account the fact that kinetic coefficients $\eta$, $\tau_\Pi$ and
$\lambda_1$ in Eq.~(\ref{eq:Pi-xixi}) are functions of $\eden$,
which in a conformal theory are given by the following power laws:
\begin{equation}
  \label{eq:eta-tauPi-lambda1}
  \eta = C \eta_0\left(\frac\eden C\right)^{(d-1)/d}, \qquad
  \tau_\Pi = \tau_\Pi^0 \left(\frac\eden C\right)^{-1/d}, \qquad
  \lambda_1 = C\lambda_1^0 \left(\frac\eden C\right)^{(d-2)/d}\, ,
\end{equation}
where, for convenience, we defined constants $\eta_0$, $\tau_\Pi^0$
and $\lambda_1^0$, and we chose the constant $C$ to be the same as in
Eq.~(\ref{eq:eden-ideal-asymp}). Integrating
Eq.~(\ref{eq:eden-eqn-bj-2}) we thus find
\begin{equation}
  \label{eq:eden-tau-bj-solution}
  \frac{\eden(\tau)}{C} =\tau^{-2+\nu} - 2 \eta_0\, \tau^{-2} 
  +  
  \left[ \frac{2(d{-}1)}d \eta_0^2 - \frac{d{-}2}{d{-}1}
  \left(\eta_0\tau_\Pi^0 - 2 \lambda_1^0\,\frac{d{-}3}{d{-}2}\right)\right]
\tau^{-2-\nu}
+ \ldots\,.
\end{equation}

In Section~\ref{bf-ads} we shall match the Bjorken flow solution
in the strongly-coupled ${\cal N}=4$ SYM theory
found in \cite{Heller:2007qt} (see also \cite{Benincasa:2007tp}) 
using AdS/CFT correspondence
and determine $\lambda_1$ in this theory.

In order to compare our results to the ones obtained in 
Ref.~\cite{Heller:2007qt}, we shall write here the equations of
second-order hydrodynamics using also the alternative representation
(\ref{ISlikePi}) for $\Pi^{\xi\xi}$ in (\ref{eq:eden-eqn-bj-2}).
We obtain the following system of equations for the energy density and
the component of the viscous flow, which we define as $\Phi\equiv
-\Pi^\xi_\xi$ 
(see~\cite{Baier:2006um}; c.f. \cite{Muronga:2001zk} for $\lambda_1=0$):
\begin{eqnarray}
  \partial_\tau \eden&=&-\frac{d}{d{-}1} \frac{\eden}{\tau}
   + \frac{\Phi}{\tau}\,,\\
  \label{eq:Pi-xixi-IS}
\tau_\Pi \partial_\tau \Phi &=& \frac{2(d{-}2)}{d{-}1} \frac{\eta}{\tau}- \Phi 
  - \frac{d}{d-1} \frac{\tau_\Pi}{\tau}\, \Phi -\frac{d{-}3}{d{-}2}\,\frac{\lambda_1}{\eta^2}\, \Phi^2\, .
\end{eqnarray}
As should be expected, the asymptotics of the 
solution of this system coincides with
Eq.~(\ref{eq:eden-tau-bj-solution}). Equation~(\ref{eq:Pi-xixi-IS})
is different from the one used in \cite{Heller:2007qt} by
the last two terms proportional to $\tau_\Pi$ and $\lambda_1$.


\section{Second-order hydrodynamics for strongly coupled ${\cal N}=4$
  supersymmetric Yang-Mills plasma}
\label{sec:SYM}

In this Section, we compute the parameters  
$\tau_\Pi$, $\kappa$, and $\lambda_1$ of the second-order hydrodynamics for 
a theory whose gravity dual is well-known:   ${\cal N}=4$ SU($N_c$) 
supersymmetric Yang-Mills theory 
in the limit $N_c\rightarrow \infty$, $g^2 N_c\rightarrow \infty$
\cite{Maldacena:1997re,Gubser:1998bc,Witten:1998qj}. 
According to the gauge/gravity duality conjecture, 
in this limit the theory at finite temperature $T$ has an effective 
description in terms of the AdS-Schwarzschild gravitational background 
with metric
\begin{equation}
ds^2_{5} = {\pi^2 T^2 L^2\over u}
\left( -f(u) dt^2 
+ dx^2 + dy^2 +dz^2 
\right) 
+
 {L^2\over 4 f(u) u^2} du^2\,,
\label{5dbg}
\end{equation}
where $f(u) = 1-u^2$, and $L$ is the AdS curvature scale \cite{Aharony:1999ti}.
The duality allows one to compute the retarded correlation functions of the 
gauge-invariant operators at finite temperature. The result of such a computation
would in principle be exact in the full microscopic theory 
(in the limit $N_c\rightarrow \infty$, $g^2 N_c\rightarrow \infty$). 
As we are interested in the hydrodynamic limit of the theory, 
here we compute the correlators in the form of low-frequency, long-wavelength 
expansions. In momentum space, the dimensionless expansion parameters 
are 
\begin{equation}
  \label{eq:wn-qn-definition}
  \wn = \frac{\omega}{2\pi T} \ll 1, 
\qquad
  \qn \equiv \frac k{2\pi T} \ll 1.
\end{equation}
Comparing these expansions to the predictions of the 
second-order hydrodynamics  
obtained in Sections \ref{kf}, \ref{ssp} and \ref{bf} for $d=4$,
we can read off the coefficients $\tau_\Pi$, $\kappa$, $\lambda_1$. 

One must be aware that the ${\cal N}=4$ SU($N_c$) supersymmetric
Yang-Mills theory posesses conserved $R$-charges, corresponding to
SO(6) global symmetry. Therefore, complete hydrodynamics of this
theory must involve additional hydrodynamic degrees of freedom --
$R$-charge densities. Our discussion of generic conformal
hydrodynamics without conserved charges can be, of course, generalized
to this case. This is beyond the scope of this paper. Here we only
need to observe that since the $R$-charge densities are not singlets
under the SO(6) they cannot contribute at {\em linear} order to the
equations for $T^{\mu\nu}$.  These contributions are therefore
irrelevant for the linearized hydrodynamics we consider in
Sections~\ref{kf},~\ref{ssp} and~\ref{sec:shear-pole}.  For the discussion
of the Bjorken flow in Section~\ref{bf} they are also irrelevant,
since (and as long as) we consider solutions with zero $R$-charge
density.

\subsection{Scalar channel}
\label{scalar_channel}

We start by computing the low-momentum expansion of the correlator
$G^R_{xy,xy}(\omega,k)$. To leading order in momentum, 
this correlation function has been previously
computed from gravity in \cite{Son:2002sd,Policastro:2002se}. 
Following \cite{Policastro:2002se}, 
here we obtain the next to leading order term in the expansion.

The relevant fluctuation of the background metric (\ref{5dbg}) is 
the component $\phi \equiv h_x^y$ of the graviton.
The retarded correlator in  momentum space is determined by the 
on-shell boundary action
\begin{equation}
S_{tot}[H_0, k] = 
\lim_{\epsilon \rightarrow 0} \Biggl( 
S_{boundary}^{grav}[ H_0,\epsilon,k]+
S_{c.t.}[ H_0,\epsilon,k]\Biggr)\,,
\end{equation}
following the prescription formulated in \cite{Son:2002sd}.
Here  $H_0(k) = H(\epsilon,k)$ is the boundary value 
(more precisely, the value at the cutoff $u=\epsilon \rightarrow 0$) of the 
solution  to the graviton's equation of motion 
(Eq.~(6.6) in  \cite{Policastro:2002se}) 
\begin{equation}
H(u,k) = H_0(k) \frac{\phi_k(u)}{
\phi_k(\epsilon)}\,.
\label{norm}
\end{equation}
A perturbative solution $\phi_k(u)$ to 
order $\wn^2$, $\qn^2$ is given by Eq.~(6.8) in \cite{Policastro:2002se}.
The gravitational action (Eq.~(6.4) in \cite{Policastro:2002se}) 
 reduces to the sum of two terms, 
the horizon contribution and the boundary contribution.
 The horizon contribution should be discarded, 
as explained in  \cite{Son:2002sd} and later justified 
in \cite{Herzog:2002pc}. The remaining boundary term, 
$S_{boundary}^{grav}[ H_0,\epsilon,k]$, is divergent in the limit $\epsilon
\rightarrow 0$, and should be supplemented by the counterterm action
 $S_{c.t.}[ H_0,\epsilon,k]$ following a procedure known as the holographic 
renormalization.\footnote{The holographic renormalization 
\cite{Skenderis:2002wp} corresponds to
 the usual renormalization of the composite operators in the dual CFT.}
In the case of gravitational fluctuations, the  counterterm action is 
\cite{Papadimitriou:2004ap} 
\begin{equation}
S_{ct} = - \frac{3 N_c^2}{4 \pi^2 L^4} 
\int\limits_{u=\epsilon} d^4 x \sqrt{-\gamma}\Biggl(
1 + \frac{L^2}{2} P - 
\frac{L^4}{12} \left( P^{kl}P_{kl} - P^2\right)\, \log{\epsilon} \Biggr)\,,
\label{gravctar}
\end{equation}
where  $\gamma_{ij}$ is the metric  (\ref{5dbg}) restricted to  $u=\epsilon$, and
\begin{equation}
P = \gamma^{ij} P_{ij}\,, \qquad
P_{ij} = \frac{1}{2} \Biggl( R_{ij} -  \frac{1}{6}
R \gamma_{ij}\Biggr)\,.
\end{equation}
Evaluating (\ref{gravctar}), we find the total boundary 
action\footnote{Terms 
quadratic in $H$ in Eq.~(\ref{trueqnewx}) should be understood as products
$H(-\omega,-k) H(\omega,k)$, and an integration
 over $\omega$ and $q$ is implied.}  
\begin{equation}
S_{tot} = - \frac{\pi^2 N_c^2 T^4}{8} \Biggl( V_4 -  
 \frac{H(\epsilon) H'(\epsilon)}{\epsilon} + 
\frac{H^2(\epsilon)}{2} -
\frac{(\qn^2-\wn^2) H^2(\epsilon)}{\epsilon}\Biggr) + 
  O(\wn^3, \wn \qn^2) + O(\epsilon)\,.
\label{trueqnewx}
\end{equation}
The boundary action (\ref{trueqnewx}) is finite in the limit 
$\epsilon \rightarrow 0$. Its fluctuation-independent part is 
$S_{tot}^0 = - P V_4$, where $P = \pi^2 N_c^2 T^4/8$ is the pressure
in ${\cal N}=4$ SYM, $V_4$ is the four-volume.
The part quadratic in fluctuations gives the two-point function.
Substituting the solution  (\ref{norm}) into Eq.~(\ref{trueqnewx})
and using the recipe of  \cite{Son:2002sd}, we find
\begin{equation}
G_{xy,xy}^R =  - \frac{\pi^2 N_c^2 T^4}{4}  \Biggl[
i \wn -\wn^2 +\qn^2 + \wn^2 \ln{2} -\frac{1}{2} \Biggr]
+ O(\wn^3, \wn \qn^2)\,.
\label{resull}
\end{equation}
Comparing  Eq.~(\ref{resull}) to the
 hydrodynamic result (\ref{hydroxyxycorr}) we obtain 
 the pressure~\cite{Gubser:1996de}, the viscosity~\cite{Policastro:2001yc} 
and the two parameters 
of the second-order hydrodynamics for  ${\cal N}=4$ SYM:
\begin{equation}\label{tauPisound}
  P = \frac{\pi^2}8 N_c^2 T^4, \quad
  \eta = \frac{\pi}8 N_c^2 T^3, \, \quad
 \tau_\Pi = \frac{2-\ln 2}{2\pi T}\,, \quad
  \kappa = 
  \frac\eta{\pi T}\,.
\end{equation}

\subsection{Shear channel} 
\label{shear_channel}

The dispersion relation (\ref{dispshear1}) manifests itself as a pole
in the retarded Green's functions $G^R_{ty,ty}$, 
 $G^R_{ty,xy}$,  $G^R_{xy,xy}$ in the hydrodynamic approximation.
To quadratic order in $k$ this dispersion relation was computed from 
dual gravity in Section 6.2 of Ref.~\cite{Policastro:2002se}. 
Here we extend that calculation to quartic order in $k$. 
This amounts to solving the differential equation for 
the gravitational fluctuation $G(u)$ \cite{Policastro:2002se}
\begin{equation}
G'' - \left( {2u\over f} - {i\wn\over 1-u}\right) G'
 + \frac1f \biggl( 2 
+ {i \wn\over 2 } - {\qn^2\over u}
+ {\wn^2 [ 4 - u(1+u)^2]\over 4 u f} \biggr) G = 0\,
\label{Gequation}
\end{equation}
perturbatively in $\wn$ and $\qn$ 
assuming $\wn \sim \qn^2$.
The solution $G(u)$ is supposed to be regular at $u=1$ 
 \cite{Policastro:2002se}.
Such a solution is readily found by writing 
\begin{equation}
G(u) = G_0(u) + \wn G_1(u) + \qn^2 G_2(u) + 
\wn^2 G_3(u) + \wn\qn^2  G_4(u) + \qn^4 G_5(u)+\cdots
\label{pertsolG}
\end{equation}
and computing the functions $G_i(u)$ 
perturbatively\footnote{Note that, for $u$ real, $G^*(u,-\wn)=G(u,\wn)$. 
This implies $\Im G_{0,2,3,5}=0$, $\Re G_{1,4}=0$.}.
The  functions $G_i(u)$ are given explicitly in Appendix \ref{append}.
To obtain the dispersion relation, one has to substitute the solution 
$G(u)$ into the equation (6.13b) of \cite{Policastro:2002se}
and take the limit $u\rightarrow 0$. The resulting equation for $\wn$,
\begin{equation}
\qn^4 + 2 \qn^2 - 4 i \wn - i \wn \qn^2\ln 2 + 2 \wn^2\ln 2 =0\,,
\label{dispfr}
\end{equation}
has two solutions one of which is incompatible with the assumption
$\wn\ll 1$. The second solution is 
\begin{equation}
\wn = -\frac{i \qn^2}{2} - \frac{i (1-\ln 2) \qn^4}{4} + O(\qn^6)\,.
\label{dispgrav}
\end{equation}
If we naively compare Eqs.~(\ref{dispshear1}), (\ref{dispgrav}), we
would get $\tau_\Pi = (1-\ln 2)/(2 \pi T)$, which is inconsistent with
the value obtained from the Kubo's formula, Eq.~(\ref{tauPisound}).
As explained in Section~\ref{sec:shear-pole}, 
this happens because the ${\cal O}(k^4)$ term in the shear
dispersion relation is fully captured only in third-order
hydrodynamics.  In other words, we confirm that Eq.~(\ref{dispshear1})
has an error at order ${\cal O}(k^4)$.

\subsection{Sound channel}

\begin{figure}[h]
\begin{center}
\epsfig{file=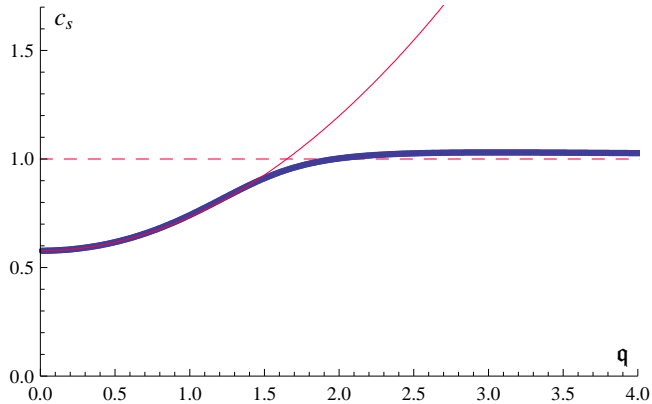,height=32ex}
\end{center}
\caption[pp1]{Sound dispersion $c_s=c_s(\qn)$ 
in  ${\cal N}=4$ SYM plasma. The dark (blue) curve 
shows the sound speed dependence on wavevector,
 $c_s(\qn) = \mbox{Re}\, \wn/ \qn$, 
with $c_s(0)=1/\sqrt{3}$ (this plot is based on 
numerical data first obtained in \cite{Kovtun:2005ev}).
 The light (red) curve
corresponds to analytic approximation derived from
Eq.~(\ref{sound_d_rel}) and valid for sufficiently small $\qn$.}

\label{sound_disp}
\end{figure}

The sound wave dispersion relations (\ref{dispsound1})
appear  as poles in the
 correlators of the diagonal components of the stress-energy tensor 
in the hydrodynamic approximation. These correlators and the 
dispersion relation to quadratic order in spatial momentum 
were first computed from gravity in \cite{Policastro:2002tn}.
A convenient method of studying the sound channel correlators was 
introduced in \cite{Kovtun:2005ev}.
 In this approach, the hydrodynamic dispersion relation
emerges as the lowest quasinormal frequency of a gauge-invariant
gravitational perturbation of the background (\ref{5dbg}).
According to  \cite{Kovtun:2005ev}, 
the sound wave pole is determined by solving the differential equation
\begin{eqnarray}
Z'' &-& {3\wn^2 (1+u^2) + 
\qn^2 ( 2u^2 - 3 u^4 -3)\over u f (3 \wn^2 +\qn^2 (u^2-3))}
 \, Z'\nonumber \\
&+& 
{3 \wn^4 +\qn^4 ( 3-4 u^2 + u^4) + \qn^2 ( 4 u^5 - 4 u^3 + 4 u^2
 \wn^2 - 6 \wn^2)\over u f^2 ( 3 \wn^2 + \qn^2 (u^2 -3))}\, Z = 0 
\label{Zequation}
\end{eqnarray}
with the incoming wave boundary condition at the horizon ($u=1$) 
and Dirichlet boundary condition $Z(0)=0$ at the boundary $u=0$,
 and taking the
 lowest frequency in the resulting quasinormal spectrum.
The exponents of the equation (\ref{Zequation}) at  $u=1$
are  $\pm i\wn/2$.
The incoming wave boundary condition is implemented by
 choosing the exponent $-i\wn/2$ and writing 
\begin{equation}
Z(u) = f^{-i\wn/2} X(u)\,,
\end{equation}
where $X(u)$ is regular at $u=1$.
Thus we obtain the following differential equation for $X(u)$
\begin{eqnarray}
X'' &+& \left( \frac{  2 u \, i \wn}{f} - \frac{1+u^2}{u f} -
\frac{4 \qn^2 \, u}{3\wn^2 + \qn^2 (u^2-3)} \right)\, X' \nonumber \\
&+& 
\Biggl( \frac{(1+u+u^2) \wn^2}{u (1+u) f} - \frac{\qn^2}{u f} -
 \frac{4 \qn^2 \, u^3 (1+i \wn) }{ u f
 (3 \wn^2 + \qn^2 (u^2-3)) }\Biggr)  X = 0\,.
\label{Xequation}
\end{eqnarray}
This equation can be solved perturbatively in $\wn\ll 1$, $\qn \ll1$ 
assuming $\wn \sim \qn$ (the expected scaling in the sound wave
 dispersion relation). Rescaling $\wn \rightarrow \lambda \wn$, 
 $\qn \rightarrow \lambda \qn$, where $\lambda \ll 1$, 
we look for a solution in the form
\begin{equation}
X(u) = X_0 (u) + \lambda X_1 (u) + \lambda^2 X_2 (u) + \cdots\,.
\label{pertsolX}
\end{equation}
The functions $X_i(u)$ are written explicitly in Appendix \ref{append}.
The Dirichlet condition $X(0)=0$ leads to the equation for $\wn(\qn)$:
\begin{eqnarray}
&-& i \wn \qn^2 + \frac{\qn^2}{2} - \frac{3 \wn^2}{2} +
\frac{\wn^4}{16} \left( \pi^2 - 12 \ln^2 2 + 24 \ln 2\right) -
\frac{\qn^4}{12} \left( 2 \ln 2 - 8\right) \nonumber \\
&-& \frac{\wn^2\qn^2}{48} 
 \left( \pi^2 - 12 \ln^2 2 + 48 \ln 2\right)=0\,.
\end{eqnarray}
To order $\qn^3$, the solution is given by 
\begin{equation}
\wn = \pm \frac{\qn}{\sqrt{3}} - 
\frac{i \qn^2}{3} \pm \frac{(3 - 2\log{2})\qn^3}{6 \sqrt{3}} +  O(\qn^4)\,.
\label{sound_d_rel}
\end{equation}
This is the dispersion relation for the sound waves to order  $\qn^3$.
The complete dispersion relation can be obtained by solving 
 the equation (\ref{Zequation}) numerically \cite{Kovtun:2005ev}.
The sound dispersion curve is shown in 
Fig.~\ref{sound_disp}.
Comparing Eq.~(\ref{sound_d_rel}) to 
Eq.~(\ref{dispsound1}) we find the relaxation time $\tau_\Pi$ for the strongly 
coupled  ${\cal N}=4$ SYM plasma:
\beq
\tau_\Pi=\frac{2- \ln2}{2 \pi T}\,.
\label{relax}
\eeq
The result (\ref{relax}) coincides with the one obtained in Section
\ref{scalar_channel}, which is a nontrivial check of our approach.

\subsection{Bjorken flow}
\label{bf-ads}

In order to determine $\lambda_1$, we match
Eq.~(\ref{eq:eden-tau-bj-solution}) with the solution found by Heller
and Janik~\cite{Heller:2007qt} given by\footnote{The quantities in
  Eq.~(\ref{eq:janik-heller}) can be thought of as dimensionless
  combinations of quantities in Eq.~(\ref{eq:eden-tau-bj-solution})
  with an appropriate power of an arbitrary scale parameter $\tau_0$:
  $\tau/\tau_0$, $\eden\tau_0^d$, $\eta_0\tau_0^{\nu}$,
  $C\tau_0^{d\nu}$ etc. Due to conformal invariance, a rescaled
  solution is also a solution, and the scale $\tau_0$ can be used
  instead of the integration constant $C$, to parameterize the
  solutions in Eq.~(\ref{eq:eden-tau-bj-solution}).}
\begin{equation}\label{eq:janik-heller}
 \eden(\tau)=\frac{N_c^2}{2\pi^2}\left[
 \tau^{-4/3}-2 \eta_0 \tau^{-2}+\tau^{-8/3} 
 \left(\frac{10}{3} \eta_0^2+ \frac{6 \ln
     2-17}{36\sqrt{3}}\right)\right],
\quad\textrm{with}\quad \eta_0=\frac{1}{\sqrt{2}\, 3^{3/4}}\,.
\end{equation}
Matching by using $C=N_c^2/(2\pi^2)$,
and $\tau_\Pi=(2-\ln 2)/(2 \pi T)$ from Eq.~(\ref{relax}), 
together with $\eden=3\pi^2N_c^2T^4/8$ 
and Eq.~(\ref{eq:eta-tauPi-lambda1}) 
gives
\begin{equation}
\lambda_1=\frac{\eta}{2 \pi T}.  
\end{equation}
Note that Heller and Janik \cite{Heller:2007qt} found a different
value for $\tau_\Pi$ since they matched to the Israel-Stewart 
equations for hydrodynamics, 
and not the more general (nonlinear) equation~(\ref{ISlikePi}).

%

\section{Kinetic theory}
\label{sec:kinetic-theory}

Our analysis should be valid not only for the strongly coupled ${\cal
N}=4$ SYM theory, but also for all theories with conformal symmetry.
In particular, it should be valid also for weakly coupled CFT like the
SYM theory at small 't Hooft coupling, or QCD at sufficiently large
$N_f$ at the Banks-Zaks fixed point \cite{Banks:1981nn}.  In these cases, one expects that
it is possible to understand and compute the second-order transport
coefficient from kinetic theory.  We set $d=4$ in this Section.

\subsection{Setup}

Since we are to discuss conformal transformations,
our starting point is the classical Boltzmann equation
in curved rather than flat space-time \cite{Stewart71,Brandt:1994mv},
\beq
\left[p^\mu \frac{\partial}{\partial x^\mu}-\Gamma^{\lambda}_{\mu \nu} p^\mu p^\nu 
\frac{\partial}{\partial p^\lambda}\right]
f(p,x)=-{\mathcal C}[f],
\label{BE}
\eeq
where $f(p,x)$ is the one-particle distribution function,
$p^\mu$ is the particle momentum, $\Gamma^{\lambda}_{\mu \nu}$
are the Christoffel symbols and ${\mathcal C}$ is the collision integral.
One can easily show that conformal transformations are a symmetry
of the Boltzmann equation if particles
are massless ($p^2\equiv p^\mu p_\mu = 0$) and 
the collision integral transforms as ${\mathcal C}[\bar{f}]\rightarrow
e^{2 \omega(x)}{\mathcal C}[f]$.

Hydrodynamic equations are obtained by taking moments with
respect to the particle momentum $p^\mu$ of Eq.~(\ref{BE}).
More precisely, acting with 
$\int d\chi \equiv \int d^4 p \delta(-p^2) \theta(p^0) $, 
where $\theta$ is the step-function, on Eq.~(\ref{BE}) one obtains
\beq
\int d\chi \sqrt{-g} \left[p^\mu \frac{\partial}{\partial x^\mu}-\Gamma^{\lambda}_{\mu \nu} p^\mu p^\nu 
\frac{\partial}{\partial p^\lambda}\right]
f(p,x)=-\int d\chi \sqrt{-g}{\mathcal C}[f],
\eeq
which upon partial integration leads \cite{Brandt:1994mv} to
\beq
\nabla_\mu \int d\chi p^\mu \sqrt{-g} f(p,x) = - 
\int d\chi \sqrt{-g}{\mathcal C}[f].
\label{CC}
\eeq
We recall here that $\nabla_\mu$ is the (geometric) covariant derivative. In
theories with conserved charges or if only elastic collisions
are considered, $\int d\chi {\mathcal C}[f]=0$
and
Eq.~(\ref{CC}) becomes the
conservation of the particle current in theories with conserved charges.
Conservation of the energy-momentum tensor\footnote{
Note that sometimes $p^\mu$ is traded by the introduction of a 
``local momentum''
\cite{Bernstein} and as a consequence $T^{\mu \nu}$ would be defined without
a factor of $\sqrt{-g}$ and the form of the Boltzmann equation (\ref{BE}) 
changes.}
\beq
T^{\mu \nu}\equiv \int d\chi p^\mu p^\nu \sqrt{-g} f(p,x)
\label{EMT}
\eeq
follows from Eq.~(\ref{BE}) upon action of $\int d\chi p^\nu$
and the requirement $\int d\chi \sqrt{-g} p^\nu {\mathcal C}[f]=0$,
\beq
\nabla_\mu T^{\mu \nu}=0.
\label{EMC}
\eeq
Acting with $\int d\chi p^\nu p^\lambda$ on Eq.~(\ref{BE})
gives the first equation with non-trivial contribution from the collision 
integral \cite{Mueller},
\beq
\nabla_\mu X^{\mu \nu \lambda}= I^{\nu \lambda},
\label{TME}
\eeq
where
\bqa
X^{\mu \nu \lambda} & \equiv & 
\int d\chi p^\mu p^\nu p^\lambda \sqrt{-g} f(p,x) \label{xdef}\, ,\\
I^{\nu \lambda} &\equiv& - \int d\chi p^\nu p^\lambda \sqrt{-g} 
{\mathcal C}[f].
\label{idef}
\eqa
Similarly, an infinity of higher moment equations of the form 
\beq
\nabla_\mu X^{\mu \nu_1 \nu_2 \nu_3 \ldots}=I^{\nu_1 \nu_2 \nu_3 \ldots}
\eeq 
also follow from Eq.~(\ref{BE}). 
%
%
%

Splitting the out-of-equilibrium particle
distribution function into an equilibrium and non-equilibrium part
\beq
f(p,x)=f_{\rm eq}(p,x)\left(1+\delta f(p,x)\right),
\label{decompeq}
\eeq
one defines an equilibrium energy-momentum tensor 
\beq
T^{\mu \nu}_{\rm eq}=T^{\mu \nu}\equiv \int d\chi p^\mu p^\nu \sqrt{-g} f_{\rm eq}(p,x)\, ,
\eeq
and a non-equilibrium component 
$\Pi^{\mu \nu}=T^{\mu\nu}-T^{\mu \nu}_{\rm eq}$, which 
by construction is both symmetric and traceless.
We shall assume that the equilibrium distribution function 
$f_{\rm eq}(p,x)=f_{\rm eq}(- u(x)\cdot p/T(x))$ depends
on local temperature and velocity $T,u_\mu$, which are defined
such that the equilibrium distribution has the same energy
and momentum density as $f$ in the rest frame defined by $u_\mu$,
\beq
\int d\chi \sqrt{-g} p^\mu (u_\nu p^\nu) \left(f-f_{\rm eq}\right)=0.
\label{constraints}
\eeq
This implies that $u_\mu \Pi^{\mu \nu}=0$.

\subsection{Moment approximation}

While the full hierachy of moment
equations should correspond to the original Boltzmann equation,
it is too complicated to be treated exactly. However,
an approximate evolution equation for systems not too far from
equilibrium may be constructed. 
The approximation is similar to the Grad's 14-moment method
\cite{Grad}.

We decompose $\delta f$ into spherical harmonics,
\beq
\delta f=\sum_{l=0}^{\infty} f^{(l)}_{\mu_1 \ldots \mu_l}(\xi)
p^{\mu_1} \ldots p^{\mu_l}, \quad \xi=-\frac{u \cdot p}{T},
\eeq
where $f^{(l)}_{\mu_1 \ldots \mu_l}(\xi)$ are fully symmetric,
orthogonal to $u^\mu$, and traceless over any pair of indices.
By construction, the $l=0,1$ parts satisfy the constraints 
Eq.~(\ref{constraints}). The approximation is now specified by the following assumptions
(c.f. \cite{deGroot}):
\begin{itemize}
\item
the system is sufficiently close to equilibrium that the
collision term is linear in $\delta f$
\item
all contributions $l>2$ are subdominant
\item
for $l\le 2$ and expanding in some basis, 
all $\xi$ dependent terms are subdominant.
\end{itemize}
This implies that 
\beq
\delta f(p,x)\sim T^{-6} p^\mu p^\nu \Pi_{\mu \nu}+{\mathcal O}(\Pi^2),
\label{phiansatz}
\eeq
and 
\beq
I^{<\nu \lambda>}\sim T^2(x) \Pi^{<\nu \lambda>}(x) +{\mathcal O}(\Pi^2),
\label{collterm}
\eeq
where subdominant terms have been labelled as ${\mathcal O}(\Pi^2)$.
It would be interesting to use numerical 
techniques such as in Ref.\cite{Arnold:2003zc,Gombeaud:2007ub} 
to test the correctness of Eq.~(\ref{phiansatz}).

Splitting $X^{\mu \nu \lambda}$ into an equilibrium and non-equilibrium
part, one finds
\beq
X^{\mu \nu \lambda}_{\rm eq}=\int d\chi p^\mu p^\nu p^\lambda \sqrt{-g} f_{\rm eq}(p,x) \sim
T^5\left[\ u^\mu u^\nu u^\lambda+
{\rm const}\times \left(\Delta^{\mu \nu} u^\lambda+perm.\right)\right],
\eeq
where $perm.$ denotes all non-trivial permutations of indices,
and
\beq
X^{\mu \nu \lambda}-X^{\mu \nu \lambda}_{\rm eq}\sim
T \Pi^{(\mu \nu} u^{\lambda)},
\eeq
where $(\mu_1 \mu_2 \ldots \mu_n)$ denotes symmetrization
with respect to the indices $\mu_1,\mu_2,\ldots,\mu_n$.
Projection $<>$ on the moment equation (\ref{TME}) thus gives
\beq
\Pi^{\nu \lambda} + \tau_{\Pi}
\left[\Pi^{\nu \lambda} D {\rm ln} T + \Delta^{\nu}_\alpha \Delta^\lambda_\beta
D \Pi^{\alpha \beta}+\Pi^{\nu \lambda} \nabla_\mu u^\mu
+2 \Pi^{\mu < \nu} \nabla_\mu u^{\lambda>}
\right]= -\eta \sigma^{\nu \lambda}+{\mathcal O}(\Pi^2),
\label{finaleq}
\eeq
where the proportionality constants have been denoted by  $\eta$ and $\tau_\Pi$,
respectively (the ratio of these can be calculated when specifying 
$f_{\rm eq}$, c.f.\cite{Baier:2006um}).
Introducing the completely symmetric tensor 
\beq
\theta_{\mu \rho}=\frac{1}{2}
\Delta^\alpha_\mu \Delta^\beta_\rho \left(\nabla_\alpha u_\beta+\nabla_\beta
u_\alpha\right)
\eeq
one can decompose
\beq
\Pi^{\mu < \nu} \nabla_\mu u^{\lambda>}= -\Pi^{\alpha (\nu} 
\Omega^{\lambda)}_{\ \alpha}
+\Pi^{\alpha (\nu} \theta_\alpha^{\ \lambda)}-\frac{1}{3} \Pi^{\alpha \beta}
\Delta^{\nu \lambda} \theta_{\alpha \beta}.
\eeq
Rewriting
\beq
\theta_{\mu \rho}=\nabla^\perp_{<\mu} u_{\rho >}+ \frac{1}{3}
\Delta_{\mu \rho} \nabla^\perp_\gamma u^\gamma
\eeq
such that 
\beq
\Pi^{\mu <\nu} \nabla_\mu u^{\lambda>}=-\Pi^{\alpha (\nu} \Omega^{\lambda
)}_{\ \alpha}+\frac{1}{3} \Pi^{\nu \lambda} \nabla_{\gamma} u^\gamma
-\frac{\Pi^{\alpha <\nu} \Pi^{\lambda>}_{ \alpha}}{2\eta} +{\mathcal O}(\Pi^3),
\label{pisplit}
\eeq
we find 
\beq
\Pi^{\nu \lambda} = -\eta \sigma^{\nu \lambda}
-\tau_{\Pi}
\left[D \Pi^{<\nu \lambda>}+\frac{4}{3}\Pi^{\nu \lambda} (\nabla\cdot u)
\right]
+2\tau_\Pi\Pi^{\alpha (\nu} \Omega^{\lambda)}_{\ \alpha}
%
+\frac{\lambda_1}{\eta^2}\Pi^{\alpha <\nu} \Pi^{\lambda>}_\alpha
+{\mathcal O}(\Pi^3)\, ,
\label{morefinaleq}
\eeq
where
$D \ln T= -\frac{1}{3} (\nabla \cdot u) + {\mathcal O}(\Pi^2)$
has been used.

Eq.~(\ref{morefinaleq}), which was derived from kinetic theory here, 
corresponds to the more general Eq.~(\ref{ISlikePi})
 with $\lambda_2=-2 \tau_\Pi \eta$ and $\lambda_3=\kappa=0$.
Note that $\lambda_1$ contains a contribution from Eq.~(\ref{pisplit})
as well as from the collision integral Eq.~(\ref{collterm}) (see below).
What is commonly referred to as Israel-Stewart theory amounts to 
setting $\lambda_1=0$. Most of the time, also the terms involving
$\nabla\cdot u$ and the vorticity $\Omega^{\mu \nu}$ are dropped.
However, note that simply dropping terms involving $\nabla\cdot u$
ruins the conformal symmetry of the equation, and thus
 the resulting equation cannot be the correct hydrodynamic
description of the system dynamics beyond leading order.

\subsection{The structure of the collision integral}

In this subsection we study the structure of the collision 
integral Eq.~(\ref{collterm}) for a simplified model where
${\mathcal C}=(u \cdot p) \frac{f-f_{\rm eq}}{\tau_\Pi}$.
We will use a gradient expansion similar to the
Chapman-Enskog method (c.f. \cite{CE}). 

Let us decompose $f$ into
\beq
f=f_{\rm eq}(-u\cdot p/T) \left(1+f_1+f_2+\ldots\right),
\eeq
where $f_1,f_2$ represent terms of first and second order in
gradients, respectively.
Solving Eq.~(\ref{BE}) iteratively in gradients we find
\bqa
f_1&=&\frac{\tau_\Pi}{p\cdot u} \frac{f_{\rm eq}^\prime}{f_{\rm eq}}
p^\mu p^\alpha \nabla_\mu \frac{u_\alpha}{T}\, ,\nonumber\\
f_2&=&\frac{\tau_\Pi^2}{(p\cdot u)^3} 
\frac{(p\cdot u)f_{\rm eq}^{\prime\prime}+T f_{\rm eq}^\prime }{f_{\rm eq}}
p^\mu p^\nu p^\alpha p^\beta 
\nabla_\mu \left(\frac{u_\alpha}{T}\right)
\nabla_\nu \left(\frac{u_\beta}{T}\right)\nonumber\\
&&-\frac{\tau_\Pi^2}{(p\cdot u)^2} \frac{f_{\rm eq}^\prime}{f_{\rm eq}}
p^\mu p^\nu p^\alpha \nabla_\nu \nabla_\mu \frac{u_\alpha}{T}
+\frac{2 \tau_\Pi^2}{(p\cdot u)^2} \frac{f_{\rm eq}^\prime}{f_{\rm eq}}
p^\mu p^\nu p^\alpha \nabla_\mu \left(\frac{u_\alpha}{T}\right)
\nabla_\nu \ln T \, .
\label{modelgrad}
\eqa
From Eq.~(\ref{collterm}) and conformal symmetry, 
to second order in gradients 
the collision integral $I^{<\gamma \delta>}$ can contain terms
$\sigma^{<\gamma}_\lambda \sigma^{\delta> \lambda}$ and 
$D \sigma^{<\gamma \lambda>}+\frac{1}{3} \sigma^{\gamma \lambda} 
(\nabla \cdot u)$
but (in particular) not $\Omega^{\gamma \delta}$ or $R^{\gamma \delta}$ since 
these terms would involve anti-symmetrization of indices which 
is not allowed by Eq.~(\ref{modelgrad}).

This indicates that the terms involving $\kappa,\lambda_3$
in Eq.~(\ref{ISlikePi}) are not contained in the Boltzmann equation.
The Boltzmann equation is only an approximation of the
underlying quantum field theory, so it is possible
that these terms -- which are second order in gradients -- 
have been lost in this coarse-graining process.
It may be possible to compute the coefficients of these
terms for QCD in the weak-coupling regime by going beyond
the lowest order gradient expansion given in \cite{Blaizot:2001nr}.

\section{Analysis of the M\"uller-Israel-Stewart theory}
\label{section:IS}

\subsection{Causality in first order hydrodynamics}
\label{sec:causality-first-order}

It is instructive to compare the second-order conformal hydrodynamics
to the M\"uller-Israel-Stewart theory. M\"uller~\cite{Mueller1} and
independently later Israel and 
Stewart~\cite{Israel:1976tn,IS0b,Israel:1979wp}, considered how to
extend the 1st order hydrodynamics.  Their primary motivation was to
eliminate the apparent relativistic acausality of the 1st order
hydrodynamic equations. Formally, the acausality is the result of the
fact that the 1st order hydrodynamic equations are not 
hyperbolic~\cite{CourantHilbert,Israel:1979wp,HiscLind}. The problem
is most clearly seen by considering the linearized equation for a
diffusive mode (e.g., shear stress or charge diffusion), which is
first order in temporal but second in spatial derivatives. A discontinuity
in initial conditions for such a mode propagates at infinite speed. In
other words, the influence of an initial condition at a point in space
is instanteneously felt by any other point.

It should be clear, however, as emphasized, e.g., by Geroch
\cite{Geroch:1995bx,Geroch:2001xs} and others \cite{Kostadt:2000ty}
that the modes which defy causality are those which are not supposed
to be described by hydrodynamics (i.e., microscopically short
wavelengths, which is clear when one thinks about discontinuities).
Nevertheless, for numerical simulations of relativistic hydrodynamic
systems such superluminal propagation is a nuisance because in such
simulations one extrapolates hydrodynamic equations to the microscopic
scale, even though the modes, or the configurations, which are being
studied are hydrodynamic. For example, superluminal propagation makes
posing initial value problem difficult: even if the initial
hypersurface is space-like, the initial values at different points can
influence each other and an attempt to specify them independently  
leads to unacceptable singular solutions~\cite{HiscLind-instab,Kostadt:2000ty}.

Since the problem lies in the domain where the theory is not
applicable, one can safely modify the theory in this domain, without
disturbing physical predictions. This is the essence of the
solution which M\"uller and Israel proposed by extending the set of
variables. The resulting system of equations is hyperbolic.  Here we
shall write down explicitly the system of equations of Israel and
Stewart, restricting to the case of conformally invariant system
without a conserved charge that we study in this paper.

\subsection{Hydrodynamic variables and second order hydrodynamics}
\label{sec:2nd-order}

As we have already emphasized in Section~\ref{sec:1st-order-hydro} the
hydrodynamics should be viewed as a controllable expansion in
gradients of the hydrodynamic variables.  The choice of the variables,
or fields, can be aided by applying the requirement that a linearized
system of equations has solutions whose frequency vanishes in the
hydrodynamic limit, i.e., when the wave vector $\bm k$ vanishes. We call
such linearized modes the hydrodynamic modes.  Fluctuations of conserved
densities are automatically hydrodynamic because their equations are
conservation laws and constant fields ($\omega=0$, $\bm k=0$) are
trivial solutions of them. 

Hence, for a system without conserved charges the set of hydrodynamic
variables consists of the densitites of energy and momentum,
represented by 4 independent covariant variables $\eden$ and
$u^\mu$ ($u{\cdot}u=-1$).  All other quantities in hydrodynamic
description are instantaneous functions of these variables and their
derivatives, such as, e.g., $\Pi^{\mu\nu}$ (Section~\ref{sec:1st-order-hydro}).

How should one extend 1st order hydrodynamics to higher derivatives?
The systematic way, as we argued in Section~\ref{sec:1st-order-hydro}
and \ref{sec:2nd}, is to continue the expansion
(\ref{eq:tmunu-1st-order}) and add all possible terms of the second
order in derivatives, as we did in Eq.~(\ref{T2nd}).

Instead, M\"uller, Israel and Stewart take a more phenomenological
point of view.  They consider $\Pi^{\mu\nu}$ -- the viscous part of
the the momentum flow -- as a set of {\em independent} additional
variables. The equations for these variables are not given by any
exact conservation laws, but by phenomenological expansions in
the set of independent variables, which now includes also~$\Pi^{\mu\nu}$:
\begin{equation}
  \label{eq:DPimunu}
  \tau_\Pi  D \Pi^{\mu\nu} = - \Pi^{\mu\nu} - \eta\sigma^{\mu\nu}\,.
\end{equation}
The first term in Eq.~(\ref{eq:DPimunu}) has a simple intuitive
meaning: in the absence of velocity gradients ($\sigma^{\mu\nu}=0$) 
the viscous momentum flows
$\Pi^{\mu\nu}$ do not vanish instanteneously (as in
Eq.~(\ref{eq:tmunu-1st-order})), but relax to zero on a microscopic but finite
timescale $\tau_\Pi$. 
The 5 equations~(\ref{eq:DPimunu}) 
together with 4 conservation laws $\nabla_\mu T^{\mu\nu}=0$
form the system of M\"uller-Israel-Stewart 
equations for 9 variables: 
$\eden$, $u^\mu$ and $\Pi^{\mu\nu}$. (For a non-conformal system with
a conserved charge this number becomes 14.) 

In the phenomenological laws in Eq.~(\ref{eq:DPimunu})
one usually considers only terms {\em linear} in the variables
$\Pi^{\mu\nu}$ and $u^\mu$.  There is {\it a priory} no reason to
neglect nonlinear terms.  By comparing Eq. (\ref{eq:DPimunu})
with Eq.~(\ref{ISlikePi}) we see that the conformal invariance {\em requires}
presence of terms proportional to $\Pi^{\mu\nu}(\nabla{\cdot}u)$, which
are non-linear, but contain the same number of derivatives. These
terms are beyond the standard linear Israel-Stewart phenomenological
theory. In addition, bilinear terms proportional to $\lambda_i$ are
also allowed to the same order in derivatives. Such terms are relevant for
simulations of the strongly coupled quark-gluon plasma in heavy ion
collisions.

The term proportional to
$\kappa$, which vanishes in flat space, has not been considered by
Israel and Stewart but, as we have seen, is necessary to determine the
correlation functions of stress-energy tensor.

Note that in this scheme both $\Pi^{\mu\nu}$ and $\sigma^{\mu\nu}$ are of
the same, i.e., first order in the expansion around equilibrium. The
term $D\Pi^{\mu\nu}$ contains one more derivative compared to
$\Pi^{\mu\nu}$ and is thus of the second order. Without loss of
precision, to second order, one can trade $D\Pi^{\mu\nu}$ for
$-D(\eta\sigma^{\mu\nu})$ or vice versa.  Similar
substitutions can be made in other second-order terms we found, as we
did when going from Eq.~(\ref{T2nd}) to Eq.~(\ref{ISlikePi}).
Therefore, within their precision, equations of Israel-Stewart
(\ref{eq:DPimunu}) 
(or, in general nonlinear case,
Eq.~(\ref{ISlikePi})) give the same result as
the systematic expansion in derivatives.

\subsection{Causality and the domain of applicability}
\label{sec:causality-is}

The attractive feature of introducing new variables is that the
resulting equations are now first order in derivatives and, most
importantly, they are hyperbolic. This means that discontinuities propagate with
finite velocities even in the shear channel. For the shear channel
this velocity (i.e., the characteristic
velocity~\cite{CourantHilbert,Stewart_prsl,HiscLind}) can be easily
obtained from the dispersion relation (\ref{dispsheareq}) by taking
$k\to\infty$:
\begin{equation}
  \label{eq:v_disc}
  v_{\rm disc}= \sqrt\frac{\eta}{\tau_\Pi\,(\eden+P)}.
\end{equation}

Although the Israel-Stewart system of equations (\ref{eq:DPimunu}) 
or our equations (\ref{ISlikePi}),
have attractive features from the point of view of the mathematical
formulation, and are especially suitable to, e.g., numerical simulations,
 care should be taken attributing physical significance to
this fact. The domain of applicability of these equations is still
the hydrodynamic domain: $\omega$, $k$ must be small compared to
microscopic scales. The second order hydrodynamic equations increase the
precision compared with the first order equations, but only if we stay
within the hydrodynamic domain.

In practice, it is convenient to use equations which are
mathematically well-behaved even where they lose physical
significance. However, care should be taken when examining the
solutions by always considering only their features in hydrodynamic
domain -- slow and long-wavelength modes. In particular, the velocity
in Eq.~(\ref{eq:v_disc}) does not correspond to any physical
propagation. Similarly, the superluminal propagation which one
recovers according to Eq.~(\ref{eq:v_disc}) in the first order theory
when $\tau_\Pi\to0$ is the result of extrapolating the theory outside
 the hydrodynamic domain.

Nevertheless it is worthwhile to note that, with the value of
$\tau_\Pi$ in strongly coupled ${\cal N}=4$ SYM that we
find in Eq.~(\ref{tauPisound}), the characteristic velocity (\ref{eq:v_disc})
equals $1/\sqrt{2(2-\log 2)}=0.6\ldots$, i.e., less than the velocity of
light.  Therefore, the system of second order equations we wrote down
can be used in, e.g., numerical simulations without additional
modifications often needed to ensure relativistic causality and
prevent occurence of singular solutions.

\subsection{Entropy and the second law of thermodynamics}
\label{sec:Entropy}

Let us consider the question of how the second law of
thermodynamics is obeyed by the second order hydrodynamics.
For that purpose take the projection of the energy-momentum conservation
equation on $u^\nu$:
\begin{equation}
  \label{eq:u-projection}
  0=-u_\nu \nabla_\mu T^{\mu\nu} = D\eden + (\eden+P)\nabla{\cdot}u +
  \Pi^{\mu\nu}\nabla_\mu u_\nu,
\end{equation}
where we used definition Eq.~(\ref{eq:Pimunu-def}), $u{\cdot}u=-1$
 and $u_\nu\Pi^{\mu\nu}=0$. For a system without a conserved charge,
the thermodynamic entropy density $s$ is a function of the energy
density such that $ds=d\eden/T$, and it also obeys $sT=\eden+P$.
Thus, Eq.~(\ref{eq:u-projection}) can be writen as
\begin{equation}
  \label{eq:ds-Pi-nabla-u}
  T\nabla_\mu(s u^\mu)= -  \Pi_{\mu\nu}\nabla^\mu u^\nu.
\end{equation}
Since $s$ is the entropy in the local rest frame, equation
(\ref{eq:ds-Pi-nabla-u}) expresses, in a covariant form, the
rate of entropy production in the local rest frame.

For a conformal system the tensor $\Pi_{\mu\nu}$ is traceless and one
can replace $\nabla^\mu u^\nu$ on the r.h.s. of
Eq.~(\ref{eq:ds-Pi-nabla-u})
with $\sigma^{\mu\nu}/2$. Using the first order hydrodynamic relation
(\ref{eq:Pimunu-1st-order}) one then finds
\begin{equation}
  \label{eq:ds-sigma2}
   \nabla_\mu(s u^\mu)= \frac\eta{2T}\, \sigma_{\mu\nu}\sigma^{\mu\nu} +
\textrm{(3rd order terms)}.
\end{equation}
Thus, if $\eta>0$, the entropy increases, provided the 3rd order
terms on the r.h.s. of Eq.~(\ref{eq:ds-sigma2}) are negligible
compared to the 2nd order term written out. This is always true
within the domain of validity of hydrodynamics.

M\"uller and Israel observed \cite{Mueller1,Israel:1976tn} that the third order
terms in Eq.~(\ref{eq:ds-sigma2}) in
their theory can be written as the divergence of a current. Indeed,
even a complete, conformally covariant, term proportional to
$\tau_\Pi$ in Eq.~(\ref{ISlikePi}) can be written in such a way.
Solving (\ref{ISlikePi}) for $\sigma^{\mu\nu}$ and substituting into
Eq.~(\ref{eq:ds-Pi-nabla-u}) we find
  \begin{eqnarray}
    \label{eq:ds-3rdorder}
  \nabla_\mu(s u^\mu)&=& \frac1{2\eta T}\, \Pi_{\mu\nu}\Pi^{\mu\nu} 
+
\nabla_\mu
\left(\frac{\tau_\Pi}{4\eta T}\Pi_{\alpha\beta}\Pi^{\alpha\beta}u^\mu\right)
\nonumber\\ &&
-\frac1{2\eta T}\,\Pi_{\mu\nu}\left(
\kappa\op{2}
+{\lambda_1}\op{3}
+{\lambda_2}\op{4}
+{\lambda_3}\op{5}
\right)
+ \ldots\,,
  \end{eqnarray}
where we used $\tau_\Pi/\eta={\rm const}{\cdot}T^{-d}$ and the
lowest order relation $D\ln T =-(d-1)\nabla{\cdot}u$. The ellipsis
in Eq.~(\ref{eq:ds-3rdorder}) denotes 4-th order corrections.
Therefore, defining non-equillibrium entropy as
\begin{equation}
  \label{eq:entropy-noneq}
  s_{\rm noneq}=
s - \frac{\tau_\Pi}{4\eta T}\Pi_{\alpha\beta}\Pi^{\alpha\beta}
\end{equation}
one can cancel the 3rd order term proportional to $\tau_\Pi$ in 
$\nabla_\mu(s_{\rm noneq} u^\mu)$.
The correction to the equillibrium entropy in Eq.~(\ref{eq:entropy-noneq})
has an intuititive meaning --  a
non-homogeneous state of the system, in which $\Pi_{\mu\nu}\neq0$, has smaller
entropy than the equilibrium state.

The remaining terms, such as e.g., $\kappa\Pi_{\mu\nu}\op{2}/(\eta T)$, do not
appear to be total derivatives. They are also not positive
definite. However, this fact cannot be used to conclude that, e.g.,
$\kappa$ must be zero. Our explicit AdS/CFT calculation shows that
$\kappa\neq 0$. As we discussed above, the 3rd order terms in
Eq.~(\ref{eq:ds-sigma2}) do not violate the second law of thermodynamics
if we stay within the domain of applicability of hydrodynamics.  In
this domain the 3rd order terms must be small compared to the second
order term on the r.h.s. of Eq.~(\ref{eq:ds-sigma2}), which is
positive definite.

Further detailed discussions on the issue of the local entropy current
can be found in~\cite{Loganayagam:2008is,Bhattacharyya:2008xc}.

\subsection{Additional non-hydrodynamic modes}
\label{sec:additional-nonhydro}

Another interesting consequence of introducing more variables,
{\it \`a la} M\"uller-Israel-Stewart, 
is that the number of modes, or branches of the
dispersion relation $\omega(\bm k)$ increases, as we have seen in
Sections~\ref{ssp} and~\ref{sec:shear-pole}. As should be expected, 
the additional poles are not hydrodynamic: those frequencies
$\omega(\bm k)$ do not vanish as $\bm k\to0$, but remain on the
order of the microscopic scale. It should be clear from the
discussion above that the position of these poles need not be
predicted correctly by the second-order theory -- they lie
outside of the regime of its validity. 

In fact, now with the knowledge of the position of Green's function
singularities in ${\cal N}=4$ SYM at strong coupling
\cite{Kovtun:2005ev} we can say that
there are infinitely many such poles. They are given by the solutions
of equations such as (\ref{Gequation}) or (\ref{Zequation}). Only the
lowest branch $\omega(\bm k)$ can be matched by hydrodynamic theory.
To describe correctly the full Green's function one needs to introduce
infinitely many degrees of freedom -- to describe infinitely many
poles.  Any theory of finite number of degrees of freedom is a
truncation.  This truncation is controllable only for the
hydrodynamic variables, which describe the poles with frequencies
vanishing as $\bm k\to0$.  The controlling parameter is the ratio of
these frequencies to a microscopic scale, i.e., $T$ in the conformal
theory, and the precision can be, in principle, increased by
increasing the order of the expansion in this parameter.

Conceptually, let us imagine that we did succeed in writing the
infinite set of extended hydrodynamic equations for infinitely many
variables, mentioned in the previous paragraph. It is easy to realize
that in a theory with gravity dual this set will be mathematically
equivalent (in the linear regime) to differential equations
(\ref{Gequation}) or (\ref{Zequation}).  The set of infinitely many
4-dimensional fields is represented by a 5-dimensional field in these
equations.




\section{Conclusion}
\label{sec:concl}

%
%

We have determined the most general form of relativistic viscous
hydrodynamics of a conformal fluid (with no conserved charges) to
second order in gradients. We find that conformal invariance reduces
the number of allowed terms relative to more general, non-conformal,
hydrodynamics. As already known, at first order in gradients only one
kinetic coefficient, the shear viscosity $\eta$, enters the
equations. At second order we find five allowed terms with coefficients
$\tau_\Pi$ (customarily referred to as relaxation time), $\kappa$,
$\lambda_1$, $\lambda_2$ and $\lambda_3$.

The general viscous hydrodynamic equations we obtained can be matched
to AdS/CFT calculations in strongly coupled $\N=4$ supersymmetric 
 Yang-Mills
theory, and for this theory we thus determined three of the five
second-order coefficients: $\tau_\Pi$, $\kappa$ and
$\lambda_1$.  We also find that for a weakly coupled conformal plasma
describable by the Boltzmann equation, two of the coefficients
vanish. However, at least one of these coefficients, i.e., $\kappa$,
is not zero for strongly coupled $\N=4$ super Yang-Mills theory. It
would be interesting to understand how this coefficient emerges as the
approximation of the Boltzmann equation breaks down at large coupling.

We emphasized the already known fact that the equations of the 
M\"uller-Israel-Stewart theory, despite their appearance, are only
applicable in the hydrodynamic regime, where their predictions
coincide with those of the second-order gradient expansion. We also
pointed out that variants of the M\"uller-Israel-Stewart theory
used in numerical simulations of relativistic plasmas 
frequently miss terms
which are not only allowed, 
but also required for conformally
invariant theories.  If the quark-gluon plasma is approximately
conformal, then
the second-order hydrodynamic equation found in
this paper should be used instead.  One may hope that 
the values of the kinetic 
coefficients $\tau_\Pi$ and $\lambda_1$, found in ${\cal N}=4$ SYM theory,
may serve as crude estimates for their values in the strongly coupled regime
of the quark-gluon plasma.

\acknowledgments 

We would like to thank Rafael Sorkin for bringing
references \cite{Geroch:1995bx,Geroch:2001xs} to our
attention, and Gary Gibbons for discussions. 
P.R. and D.T.S. would like to acknowledge financial support by US DOE,
grant number DE-FG02-00ER41132. The work of M.A.S. is supported by the
DOE grant No.\ DE-FG0201ER41195. The work of A.O.S. is supported
by the STFC Advanced Fellowship. M.A.S. and A.O.S. would like to thank
the Isaac Newton Institute (Cambridge, U.K.) for hospitality 
during the program ``Strong Fields, Integrability and Strings,'' when
part of this work was carried out. 

{\sl Note added:} After this work was completed, we become aware of
Ref.~\cite{BHMR} where second-order hydrodynamics is derived from gravity
in AdS$_5$ space.  We thank S.~Minwalla for giving us a preview of
Ref.~\cite{BHMR}.

\newpage

\appendix

\section{Perturbative solutions of the shear and the 
sound mode equations}
\label{append}

{\bf The shear mode}

The functions $G_i(u)$ entering the perturbative solution 
(\ref{pertsolG}) of the equation (\ref{Gequation}) are 
\begin{equation}
G_0(u) = C u\,, \qquad 
G_1(u) = i C \left( u - 1 + \frac{u}{2} \ln{\frac{u+1}{2}}\right)\,, 
\qquad G_2(u) = \frac{C(1-u)}{2}\,, 
\end{equation}
\begin{eqnarray}
G_3(u) &=& - \frac{C}{48} \Biggl(
6 \pi^2 u -24 (u+1) \ln{2} - i 12 \pi u \ln 2 
- 6 u \ln^2 2 +
18 u \ln^2 (u-1) \nonumber \\ &+& 24 (u+1) \ln (u+1)
+ 12 u \ln 2 \ln{\frac{1+u}{1-u}} -12 u \ln (1+u)\ln{\frac{1+u}{1-u}}+
6 u \ln^2\frac{1+u}{1-u} \nonumber \\&-& 24 u \mbox{Li}_2 \left( \frac{1-u}{2}\right) + 12 u \ln (u-1) \Biggl( \ln 2 - 2\ln (1-u)
 - i \pi \Biggr)
\Biggr)\,, \nonumber \\
G_4(u) &=&  \frac{C}{16} \Biggl( - 4 \pi u - 4 i (1+ 3 u) \ln 2
+ 4 i \ln (1+u) + 16 i u \ln{\frac{1+u}{u}} 
+ 2 i u \ln (u-1)\left( \ln{\frac{1+u}{1-u}}\right)
 \nonumber \\
&-& 4 i u \ln{\frac{1+u}{1-u}}+ 2 \pi u \ln{\frac{1+u}{1-u}} - 2 i u \ln (1+u)
\ln{\frac{1+u}{1-u}} + 2 i u \ln^2 \frac{1+u}{1-u} - 4 i u \ln (u-1) \Biggr)\,, \nonumber \\
G_5(u) &=&  \frac{C}{4}  \Biggl( 1- u - 2 u \ln{\frac{1+u}{2u}}\Biggr)\,,
\end{eqnarray}
where $C$ is a constant, $\mbox{Li}_2(z)$ is a polylogarithm.

An alternative
way to obtain the dispersion relation (\ref{dispfr}) 
is the following:
the functions $G_i(u), i = 0,1, ..5 $ satisfy the 
inhomogeneous differential equations
\beq
(1 - u^2) G_i^{''} - 2 u G_i^{'} + 2 G_i = F_i(u) ~,
\label{Legendre}
\eeq
with $F_0 = 0, F_1 = -i (1 +u) G_0^{'} - i/2 G_0, etc.$
The homogeneous part of (\ref{Legendre}) is the Legendre differental equation
with the Legendre functions $P_1(u) = u$ and $Q_1(u) =
\frac{u}{2}\ln{\frac{1+u}{1-u}} -1$ as solutions.
Therefore $G_0 = C u$, and
for $i \ge 1$
\beq
G_i(u) = P_1(u) \int_{u}^1~ Q_1 (u') F_i (u') du'
 - Q_1(u) \int_{u}^1 ~P_1(u')F_i(u')du' ~,
\label{gsol}
\eeq
regular at $u = 1$.
Finally, the values at $u = 0$ are obtained by
\beq
G_i(u = 0) = \int_{0}^1 ~ u F_i(u) du~,
\eeq
i.e. $G_0(0)= 0, G_1(0) = -iC, $
and
\begin{eqnarray}
G_2(0) &=& \int_0^1 G_0(u) du = C/2, \nonumber \\
G_3(0) &=& \frac{C}{4} \int_0^1 ~u \left[ (2 +3 u) \ln{\frac{1+u}{2}}
+ 7 u - \frac{2}{1+u} \right] du = C \frac{\ln 2}{2} \, ,
\label{coeff}
\end{eqnarray}
etc., and hence we find Eq.~(\ref{dispfr}).

\noindent
{\bf The sound mode}

The functions $X_i(u)$ of the  perturbative solution 
(\ref{pertsolX}) of the equation (\ref{Xequation}) are 
\begin{equation}
X_0 (u) = \frac{(\qn^2 + \qn^2 u^2 - 3 \wn^2) C}{4 \qn^2}\,,
\qquad X_1 (u) = - \frac{i C \wn f(u)}{2}\,,
\end{equation}
\begin{eqnarray}
X_2 (u) &=& \frac{C}{48 \qn^2} \Biggl[ 
2 \qn^2  \Biggl( 8 - 8 u - i \pi (1+u^2) -
(1+u^2) \, 2 \ln{2} \Biggr) \nonumber \\  &+&
 3 \wn^4 \Biggl( \pi^2 - 6 i \pi - \ln{8} \left( \ln{8} - 4\right)\Biggr)
 \nonumber \\ &+&
 \qn^2 \wn^2 \Biggl( 6 i \pi (2+u^2) - \pi^2 (1+u^2)
- 24\,  ( u^2 - u + \ln{2} ) +
  \ln{8} \left(  \ln{8} +u^2 (4 + \ln{8})\right)\Biggr)\nonumber \\
&-& 2 (\qn^2 - 3 \wn^2 ) ( \qn^2 (1+u^2) - 3 \wn^2) 
\left( - i \pi + \log{(1-u)}\right)\nonumber \\
&+& 4 \Biggl( \qn^4 (1+u^2) + 9 \wn^4 (\ln{2}-1) 
 - 3 \wn^2 \qn^2
 \left( \ln{2}-2 + u^2 (\ln{2}+1)\right)\Biggr) \ln{(1+u)} 
\nonumber \\ &+& 3 \wn^2 \left( \qn^2 (1+u^2) - 3 \wn^2\right)
 \ln^2 (1+u)\nonumber \\
&+& 2 \left( \qn^2 (1+u^2) - 3 \wn^2\right)
\left( \qn^2 - 3 \wn^2 ( 1 + \ln{2}) + 3 \wn^2 \ln{(1+u)} \right) \ln{(1-u)}
\nonumber \\
 &+& 6 \wn^2 \left( \qn^2 (1+u^2) - 3 \wn^2 \right) 
\mbox{Li}_2\left(\frac{1+u}{2} \right)\Biggr]\,.
\end{eqnarray}


\end{document}